\documentclass{aa}
\usepackage{graphicx}
\usepackage{float}
\usepackage{amssymb}
\usepackage{amsmath}
\usepackage{color}
\usepackage{multirow}
\usepackage[T1]{fontenc}
\usepackage[utf8]{inputenc}
\usepackage[draft]{hyperref}

\newcommand{\sarc}{$^{\prime\prime}\!\!$}

\newcommand{\wphz}{$\,$W$\,$Hz$^{-1}$}

\newcommand{\civ}{C$\,$\textsc{iv}}
\newcommand{\mgline}{Mg$\,$\textsc{ii}~2800\AA}
\newcommand{\alline}{Al$\,$\textsc{iii}~1857\AA}
\newcommand{\zrange}{$1.7\leq z\leq 4.3$}

\newcommand{\logr}{log($R_{\textrm{144 MHz}}$)}
\newcommand{\logrstar}{log($R_{\textrm{5 GHz}}$)}
\newcommand{\llofar}{$L_{\textrm{144 MHz}}$}
\newcommand{\lfirst}{$L_{\textrm{1.4 GHz}}$}
\definecolor{Mygrey}{gray}{0.75}

\begin{document}
\title{The origin of radio emission in broad absorption line quasars: Results from the LOFAR Two-metre Sky Survey\thanks{Catalog is available in electronic form
at the CDS via anonymous ftp to \href{cdsarc.u-strasbg.fr}{cdsarc.u-strasbg.fr} (130.79.128.5)
or via \href{http://cdsweb.u-strasbg.fr/cgi-bin/qcat?J/A+A/}{http://cdsweb.u-strasbg.fr/cgi-bin/qcat?J/A+A/}} }
\titlerunning{BALQSOs in LoTSS DR1}
\author{L. K. Morabito,\inst{1}\thanks{E-mail: leah.morabito@physics.ox.ac.uk} 
\and J. H. Matthews\inst{1} 
\and P. N. Best\inst{2} 
\and G. G\"{u}rkan\inst{3} 
\and M. J. Jarvis\inst{1,4}
\and I. Prandoni\inst{5} 
\and K. J. Duncan\inst{6}
\and M. J. Hardcastle\inst{7} 
\and M. Kunert-Bajraszewska\inst{8} 
\and A. P. Mechev\inst{6}
\and S. Mooney\inst{9} 
\and J. Sabater\inst{2} 
\and H. J. A. R\"{o}ttgering\inst{6}
\and T. W. Shimwell\inst{6,10}
\and D. J. B. Smith\inst{7} 
\and C. Tasse\inst{11,12}
\and W. L. Williams\inst{7}
} 
\authorrunning{L.K. Morabito et al.}
\institute{$^{1}$Astrophysics, University of Oxford, Denys Wilkinson Building, Keble Road, Oxford, OX1 3RH, UK \\ 
$^{2}$SUPA, Institute for Astronomy, Royal Observatory, Blackford Hill, Edinburgh, EH9 3HJ, UK \\ 
$^{3}$CSIRO Astronomy and Space Science, PO Box 1130, Bentley WA 6102, Australia \\
$^{4}$Department of Physics, University of the Western Cape, Cape Town 7535, South Africa \\
$^{5}$INAF- Istituto di Radioastronomia, via P. Gobetti 101, 40129 Bologna. Italy \\
$^{6}$Leiden Observatory, Leiden University, P.O. Box 9513, NL-2300 RA, Leiden, The Netherlands \\
$^{7}$Centre for Astrophysics Research, School of Physics, Astronomy and Mathematics, University of Hertfordshire, College Lane, Hatfield AL10 9AB, UK \\
$^{8}$Toru\'n Centre for Astronomy, Faculty of Physics, Astronomy and Informatics, NCU, Grudziacka 5, 87-100 Toru\'n, Poland \\
$^{9}$School of Physics, University College Dublin, Belfield, Dublin 4, Republic of Ireland \\
$^{10}$ASTRON, The Netherlands Institute for Radio Astronomy, Postbus 2, 7990 AA, Dwingeloo, The Netherlands \\
$^{11}$GEPI \& USN, Observatoire de Paris, Université PSL, CNRS, 5 Place Jules Janssen, 92190 Meudon, France \\
$^{12}$Department of Physics \& Electronics, Rhodes University, PO Box 94, Grahamstown, 6140, South Africa } 
\date{Received XXX; accepted YYY}
\abstract{We present a study of the low-frequency radio properties of broad absorption line quasars (BALQSOs) from the LOFAR Two-metre Sky-Survey Data Release 1 (LDR1). The value-added LDR1 catalogue contains Pan-STARRS counterparts, which we match with the Sloan Digital Sky Survey (SDSS) DR7 and DR12 quasar catalogues. We find that BALQSOs are twice as likely to be detected at 144$\,$MHz than their non-BAL counterparts, and BALQSOs with low-ionisation species present in their spectra are three times more likely to be detected than those with only high-ionisation species. The BALQSO fraction at 144$\,$MHz is constant with increasing radio luminosity, which is inconsistent with previous results at 1.4$\,$GHz, indicating that observations at the different frequencies may be tracing different sources of radio emission. We cross-match radio sources between the Faint Images of the Radio Sky at Twenty Centimeters (FIRST) survey and LDR1, which provides a bridge via the LDR1 Pan-STARRS counterparts to identify BALQSOs in SDSS. Consequently we expand the sample of BALQSOs detected in FIRST by a factor of three. The LDR1-detected BALQSOs in our sample are almost exclusively radio-quiet (\logr $\,<2$), with radio sizes at 144$\,$MHz typically less than $200\,$kpc; these radio sizes tend to be larger than those at 1.4$\,$GHz, suggesting more extended radio emission at low frequencies. We find that although the radio detection fraction increases with increasing balnicity index (BI), there is no correlation between BI and either low-frequency radio power or radio-loudness. This suggests that both radio emission and BI may be linked to the same underlying process, but are spatially distinct phenomena.}
\maketitle

\section{Introduction}
Quasars and active galactic nuclei (AGN) often produce powerful outflows as they accrete material from their host galaxies. These outflows can take the form of collimated radio jets or wider-angle winds emanating from the accretion disc. Some of the most striking evidence for winds comes from broad absorption line quasars \citep[BALQSOs;][]{foltz_complex_1987,weymann_comparisons_1991}. BALQSOs are a subset of quasars which exhibit broad, blue-shifted absorption lines in their ultraviolet (UV) spectra, providing clear evidence of outflowing material intersecting the line of sight. The outflowing winds are a natural means for the AGN to provide feedback \citep[e.g.,][]{king_black_2003} impacting its host galaxy's evolutionary path.

In addition to deep, blue-shifted absorption troughs, BALQSOs are often highly reddened by dust absorption \citep{sprayberry_extinction_1992,reichard_continuum_2003} which can obscure the optical continuum. The fraction of BALQSOs in optical quasar samples is typically found to be about 10 percent \citep[e.g.,][]{weymann_comparisons_1991,trump_catalog_2006,gibson_catalog_2009}, but there is growing evidence from other wavebands that the intrinsic fraction is higher. Studies over the last couple decades have found intrinsic fractions closer to 20 percent (or higher) from the infrared \citep[e.g.,][]{dai_2mass_2008,ganguly_fraction_2008,maddox_luminous_2008}, radio \citep[e.g.,][]{becker_properties_2000}, or re-analysis of optical wavebands \citep[e.g.,][]{allen_strong_2011,knigge_intrinsic_2008,dai_intrinsic_2012}. They are intrinsically  X-ray weak \citep{sabra_pg_2001,clavel_long_2006,grupe_first_2008,morabito_unveiling_2014} as well as obscured, with typical intrinsic absorbing column densities of $N_H\sim 10^{22-24}\,$cm$^{-2}$. X-ray observations also suggest that the X-ray absorbing material may be located along a different line of sight than the UV absorbing winds \citep{morabito_suzaku_2011}. 

BALQSOs can be classified according to the absorption lines present in their spectra. Systems which show only absorption in high ionization species like C\textsc{iv} and N\textsc{v} are known as HiBALs and are more common, while a small subset ($\sim 10\%$) also show absorption in low ionization lines like \mgline\ and \alline\ (LoBALs) or even Fe\textsc{ii} and Fe\textsc{iii} lines (FeLoBALs). LoBALs tend to be found in systems with particularly broad, deep absorption troughs in C\textsc{iv} \citep{weymann_comparisons_1991,voit_low-ionization_1993, reichard_continuum_2003,filiz_ak_dependence_2014}. LoBALs and especially FeLoBALs have been posited to be at a particular evolutionary stage of BALQSOs {\citep[e.g.,][]{farrah_evidence_2007,lipari_gemini_2009,hall_implications_2011}, but can also be explained by geometric models where their lines  of sight intersect cooler material  \citep{elvis_structure_2000,dai_intrinsic_2012,matthews_testing_2016}, and are a particularly interesting sub-sample of BALQSOs. 

Unifying BALQSOs with non-BAL quasars is normally done via geometric models, in which BALQSOs are quasars viewed along a particular line of sight that intersects with outflowing material \citep[e.g.,][]{weymann_comparisons_1991,elvis_structure_2000,ghosh_physical_2007}; or evolutionary models, in which BALQSOs represent an early stage in the evolution of a quasar \citep[e.g.,][]{hazard_nine_1984,surdej_geometry_1987,lipari_gemini_2009}. In the former, the BAL phenomenon is often incorporated into a model with an equatorial disc wind that can also produce broad emission lines \citep{murray_accretion_1995,de_kool_radiation_1995,elvis_structure_2000}. Overall, however, BALQSOs and their winds must be subject to both evolutionary and viewing angle effects; quasars evolve and quasars have an orientation. This presents a major obstacle to understanding the BAL phenomenon. One simple example is the overall line opacity in the BAL trough: an increase in opacity could be indicative of a  higher mass loss rate (favouring evolution), or a sight-line that cuts through more optically thick material. This is seen in radiative transfer simulations \citep{borguet_polar+equatorial_2010,higginbottom_simple_2013,matthews_testing_2016}. The similarity in emission line properties between BALQSOs and non-BAL quasars \citep{weymann_comparisons_1991,dipompeo_rest-frame_2012,matthews_quasar_2017,yong_properties_2018} hints at a problem with simple orientation models, but this is tempered by the lack of a reliable orientation indicator in quasars generally \citep[e.g.][]{marin_are_2016}. The picture is confused further by polarisation measurements implying equatorial winds \citep{goodrich_polarization_1995,cohen_spectropolarimetry_1995,ogle_polarization_1999}, the discovery of polar winds in BALQSOs \citep{zhou_polar_2006,ghosh_physical_2007} and some inferred BAL outflow distances of $>100$pc \citep{chamberlain_large-scale_2015,chamberlain_strong_2015}.

The need for a clear indication of orientation has led previous studies to explore the radio properties of BALQSOs. The radio loudness parameter describes the ratio of radio to optical luminosity  \citep[e.g.,][]{kellermann_vla_1989}, $R = L_{\textrm{radio}} / L_{\textrm{optical}}$, and sources with higher or lower values of $R$ are referred to as radio-loud or radio-quiet, respectively. Initially BALQSOs were found in only radio-quiet sources \citep{stocke_radio_1992}, but subsequent radio observations showed that BALQSOs could indeed be radio-loud \citep{brotherton_discovery_1998,becker_properties_2000,gregg_discovery_2000}. Spectroscopic follow-up of the Faint Images of the Radio Sky at Twenty Centimeters \citep[FIRST;][]{becker_first_1995} showed a higher incidence of BALQSOs than seen in optical surveys, and established that while most BALQSOs are radio-quiet to radio-moderate, a significant number are radio loud \citep{becker_properties_2000}. In radio-loud quasars, with extended jets that can be easily resolved, observations of the jets themselves give a clear indication of the orientation \citep[e.g.,][]{barthel_is_1989,morabito_investigating_2017}. 

For radio-loud quasars, unification of quasars and BALQSOs via geometry would imply changes in the radio spectrum. This arises from the relativistic beaming of radio emission from a jet close to the line of sight \citep[e.g.,][]{orr_relativistic_1982}, which produces flat-spectrum radio emission in core-dominated sources, and steep-spectrum emission otherwise. This radio component of unification by orientation models is only useful if the radio emission does in fact arise from synchrotron-emitting jets/lobes. The radio emission mechanisms in radio-quiet AGN and quasars are still under debate, and may arise from accretion-related processes such as small-scale jets \citep[e.g.,][]{white_radio-quiet_2015} or disc-winds \citep{blundell_origin_2007}; alternatively, it could be due to star formation \citep{padovani_vla_2011}. Recent radio studies using very long baseline interferometry (VLBI) have shown that `radio-loud' BALQSOs tend to have small-scale jets with typical sizes of tens of kiloparsecs \citep{jiang_evn_2003,liu_compact_2008,bruni_parsec-scale_2013,kunert-bajraszewska_vlbi_2015,ceglowski_vlbi_2015}. There is no general consensus amongst these studies whether orientation or evolution is the dominant factor in the unification of BALQSOs and non-BAL quasars.

Several studies of BALQSOs suggest that they have similar radio properties to compact steep spectrum (CSS) or gigahertz-peaked spectrum (GPS) sources, which are generally thought to be young \citep{fanti_nature_1990}, or frustrated jets in dense environments \citep{van_breugel_studies_1984}. BALQSOs certainly have small linear sizes which are consistent with CSS/GPS sources. \cite{montenegro-montes_radio_2008} found that for a sample of 15 BALQSOs, all sources had convex spectra, most with peak frequencies between 1 and 5 GHz, typical for CSS/GPS sources. More recently, \cite{bruni_radio_2012} found that the incidence of GPS sources amongst BALQSOs was the same as that for non-BAL quasars, inconsistent with BALQSOs as a complete class of younger objects. 

The advent of the new LOFAR Two-metre Sky Survey \citep[LoTSS;][]{shimwell_lofar_2017} is an excellent opportunity to study an all-sky sample of thousands of BALQSOs with unprecedented sensitivity in the radio regime, and add information about the low-frequency radio properties. LoTSS Data Release 1 \citep[LDR1;][]{shimwell_lofar_2018} covers just over 400 square degrees and contains about 320,000 discrete radio sources. LoTSS has similar resolution to FIRST, but is about an order of magnitude deeper for sources with typical synchrotron spectra. Radio emission in extragalactic sources is generally ascribed to two processes: synchrotron and Bremsstrahlung (also called free-free). Synchrotron emission yields a power law with a spectral index such that $S\propto \nu^{-\alpha}$; the flux density $S$ will be brighter and dominate at low frequencies. Free-free emission has a fairly flat spectral index and a low frequency cutoff around 1$\,$GHz. Synchrotron processes will therefore dominate at radio frequencies below 1$\,$GHz, while radio observations above this frequency, like those from FIRST, can exhibit a combination of emission from both synchrotron and free-free emission. The low observing frequency of LoTSS -- 144$\,$MHz -- ensures that radio spectra are dominated by synchrotron emission. Combining FIRST and LoTSS detections of BALQSOs is also useful to help determine the overall radio properties of BALQSOs.

In this paper we use LDR1 to study, for the first time, the low-frequency radio properties of BALQSOs. We examine the observed low-frequency radio properties, and also how they correlate with the balnicity index (BI; a property which describes the BAL outflows, see Sect.~\ref{sec:ldr1}). We also exploit the sophisticated cross-matching of radio sources to Pan-STARRS1 \citep{chambers_pan-starrs1_2016} in LDR1 as a bridge between FIRST and SDSS to yield a significantly higher number of BALQSO/FIRST associations than previously found. This provides a more complete picture of the radio properties of BALQSOs than in previous studies. 

The paper is organised as follows. The observational data and BALQSO identification are described in Sect.~\ref{sec:obs_data}, followed by the results in Sect.~\ref{sec:results}. Sections~\ref{subsec1}--\ref{subsec6} cover general radio properties, while Sect.~\ref{subsec7} covers absorption line properties. A discussion of these results can be found in Sect.~\ref{sec:discussion}. The summary and conclusions are given in Sect.~\ref{sec:conclusions}.

Throughout this paper, we have assumed a cosmology in accordance with \cite{planck_collaboration_planck_2016}: $H_0=67.8$, $\Omega_m=0.308$, and $\Omega_{\Lambda}=0.692$. To handle small numbers when counting sources, counting errors are {\em always} estimated using Monte-Carlo simulations drawn from Poissonian distributions; in the large number limit this converges to the Gaussian distribution. Flux density $S$ is related to the spectral index $\alpha$ by $S\propto\nu^{\alpha}$. 

\section{Observational data}
\label{sec:obs_data}
\subsection{LoTSS Data Release 1: value-added catalogue}
\label{sec:ldr1}
LDR1 \citep{shimwell_lofar_2018} contains almost 320,000 unique radio sources and covers over 400 square degrees with 58 individual pointings of 8 hours each. These were mosaicked into a mostly continuous area between right ascension 161 to 231 degrees and declination 45.5 to 57 degrees. The average sensitivity of the survey is 70$\,\mu$Jy$\,$beam$^{-1}$, although this varies throughout the region due to data quality, calibration accuracy, and dynamic range limitations. The astrometry of LDR1 is tied to Pan-STARRS1 \citep[PS1;][]{chambers_pan-starrs1_2016,magnier_pan-starrs_2016}. The mean offset from both FIRST and WISE is less than 0.05 arcsec, with standard deviations from this value of 0.8 arcsec (WISE) and 1.2 arcsec (FIRST). The flux scale was bootstrapped from Very Large Array Low-frequency Sky Survey Redux \citep[VLSSr;][]{lane_vizier_2014} and Westerbork Northern Sky Survey \citep[WENSS;][]{rengelink_westerbork_1997}, following the method described in \cite{hardcastle_lofar/h-atlas:_2016}. Remaining uncertainties were estimated by comparing the flux densities in each pointing with TGSS \citep{intema_gmrt_2017}, and are generally no more than 20 percent. 

The LDR1 catalogue (version 1.2) includes value-added information from PS1 and the AllWISE Source catalogue \citep{cutri_vizier_2012}. The LDR1 radio sources were morphologically classified and then cross-matched to the optical and infrared (IR) data via a combination of likelihood ratio matching \citep[e.g.,][]{mcalpine_likelihood_2012,smith_herschel-atlas:_2011} and visual identification through the use of a LOFAR Galaxy Zoo project. For details of the morphological classification and all cross-matching, see \cite{williams_lofar_2018}. Overall 73 percent of LDR1 radio sources have matches to the optical/IR data. The value-added catalogue includes fluxes from PS1 bands $g,r,i,z,y$, and WISE bands $1,2,3,4$. Photometric redshifts exist in the value-added catalogue \citep{duncan_lofar_2018} but are not used in this paper as we use spectroscopic redshifts from the Sloan Digital Sky Survey, as discussed in the next sub-section.

\subsection{Optical data and derived quantities}
We used the quasar catalogues from the Sloan Digitial Sky Survey (SDSS) Data Release 12 \citep[DR12;][]{paris_sloan_2017} and Data Release 7 \citep[DR7;][]{shen_catalog_2011}. During the time of writing, DR14 became available but would have only increased the total sample size by 4 percent within the redshift range considered here. Weighing this small increase against the fact that BALQSOs in DR14 were identified via automated methods only, which can be inaccurate, we chose not to use DR14. The DR7 and DR12 catalogues contain SDSS sources which have been visually identified from their spectra as quasars. Within the area covered by LDR1, which we defined by generating a Multi-Order Coverage map (MOC)\footnote{The MOC was created using Aladin \citep{bonnarel_aladin_2000}.} from the LDR1 catalogue, all quasars are in both the DR12 and DR7 quasar catalogues. We took the DR12 catalogue as the basis for our sample, and supplemented this with additional data columns from DR7 where they did not exist in DR12. We added bolometric luminosities and Eddington ratios where they exist in the catalogue from \cite{shen_catalog_2011}, and filled in values from \cite{kozlowski_virial_2017} where they do not exist in the \cite{shen_catalog_2011} catalogue. For the relevant redshift range in this study, bolometric luminosities are available for over 99.8 percent of the sample. Eddington ratios are available for over 75 percent of LoBALs, over 83 percent of all BALQSOs, and 94 percent of non-BAL quasars. Considering all sources (i.e., whether or not they are LDR1 detected) increases these percentages slightly, but only by 1-3 percent.

Within the LDR1 sky coverage, there are 21$\,$812 quasars. This sample is 100 percent spectroscopically complete with the visually inspected redshifts of DR12. We classify BALQSOs using the {\em balnicity index} (BI), which was proposed by \cite{weymann_comparisons_1991} and is widely used to classify BALQSOs. The BI of a quasar is defined as
\begin{equation}
\mathrm{BI} = - \int^{3000}_{25000} 
\left[ 1 - \frac{F(v)}{0.9} \right] C_{B} dv,
\end{equation}
where $v$ is velocity with respect to line centre 
in km~s$^{-1}$ and $F(v)$ is the 
continuum-normalised flux. 
The constant $C_B$ is set to 1 once $F(v)$ has been less than
$0.9$ for at least 2000~km~s$^{-1}$, 
otherwise $C_B=0$. An object is then classified as a BALQSO if $\mathrm{BI}>0$. An alternative metric for classifying BALQSOs is the 
{\em absorption index} (AI), originally described by
\cite{hall_unusual_2002} and adapted by \cite{trump_catalog_2006}. The AI is given by 
\begin{equation}
\mathrm{AI} = - \int^{0}_{29000} 
\left[ 1 - F(v) \right]  C_{A} dv,
\end{equation}
where the 
constant $C_A$ is set to 1 in all troughs that satisfy $F(v)<0.9$ over an interval wider than 1000~km~s$^{-1}$. A number of differences exist between AI and BI selected samples \citep[see e.g.][]{knigge_intrinsic_2008}, such that it is possible to measure BI$\,=0$ for AI$\,>0$. Given these differences, and the bimodality in AI distributions \citep{knigge_intrinsic_2008}, we choose only `bona-fide' BALQSOs by requiring BI$\,>0$ and marked in the DR12 catalogue as BAL via visual identification. Selecting objects with BI$\,>0$ yields a total of 1$\,$045 BALQSOs.

To identify low-ionisation BALQSOs (LoBALs), which exhibit broad absorption lines from \mgline\ and/or \alline, we first calculated the AI in \mgline\ and \alline\  using the DR12  pipeline model fits to normalise to the continuum. We then constructed a LoBAL shortlist from our BALQSO catalogue, requiring that objects have either AI$\,>0$ or an absorption equivalent width in the DR12 pipeline model fit. This shortlist was then visually examined to confirm the presence of LoBAL absorption. Our criteria for LoBAL identification is thus that the object was already identified as a BALQSO in the SDSS DR12, and shows additional broad absorption in {\em one or both} of the \mgline\ and \alline\ lines, though we do not require that the LoBALs meet the more stringent BI criterion in these lines (only in \civ).  Such a definition is appropriate given the difficulties in accurately determining BI in low ionization lines \citep[see e.g.][]{weymann_2002,hall_unusual_2002,hewett_frequency_2003,dai_intrinsic_2012}. Classifying LoBALs in this way also means that our LoBAL selection is roughly equivalent to the `AI-LoBAL' sample of \cite{dai_intrinsic_2012}. The remaining BALQSOs we treat as HiBALs. As we use the \mgline\ and \alline\ to identify LoBALs, this necessarily limits our redshift range to where either line can be identified within the SDSS coverage. This is described in the next sub-section. 

\subsection{Quasars and BALQSOs in LDR1}
We cross-matched $\geq 5\sigma$ sources from LDR1 and the combined DR12/DR7 catalogue using the optical positions in each catalogue and a 1\sarc\ \ search radius. The final sample is limited in redshift, as BALQSOs can only reliably classified via their \civ\ troughs when they fall within the observed wavelength coverage of SDSS. We inspected the redshift distributions and fraction of BALQSOs as a function of redshift, see Figure~\ref{fig:f1}, to determine a redshift range of \zrange\ for the final sample. Below $z=1.7$, the fraction of BALQSOs drops (artificially) to virtually zero due to the wavelength coverage at that redshift. Above $z=4.3$, the fraction of BALQSOs increases slightly, but the rapidly decreasing number of total sources means that comparing BALQSOs to quasars becomes increasingly less meaningful. An upper redshift limit of $4.3$ also ensures that the wavelength range always includes \alline\ for LoBAL identification.

\begin{figure*}
\includegraphics[width=0.49\textwidth]{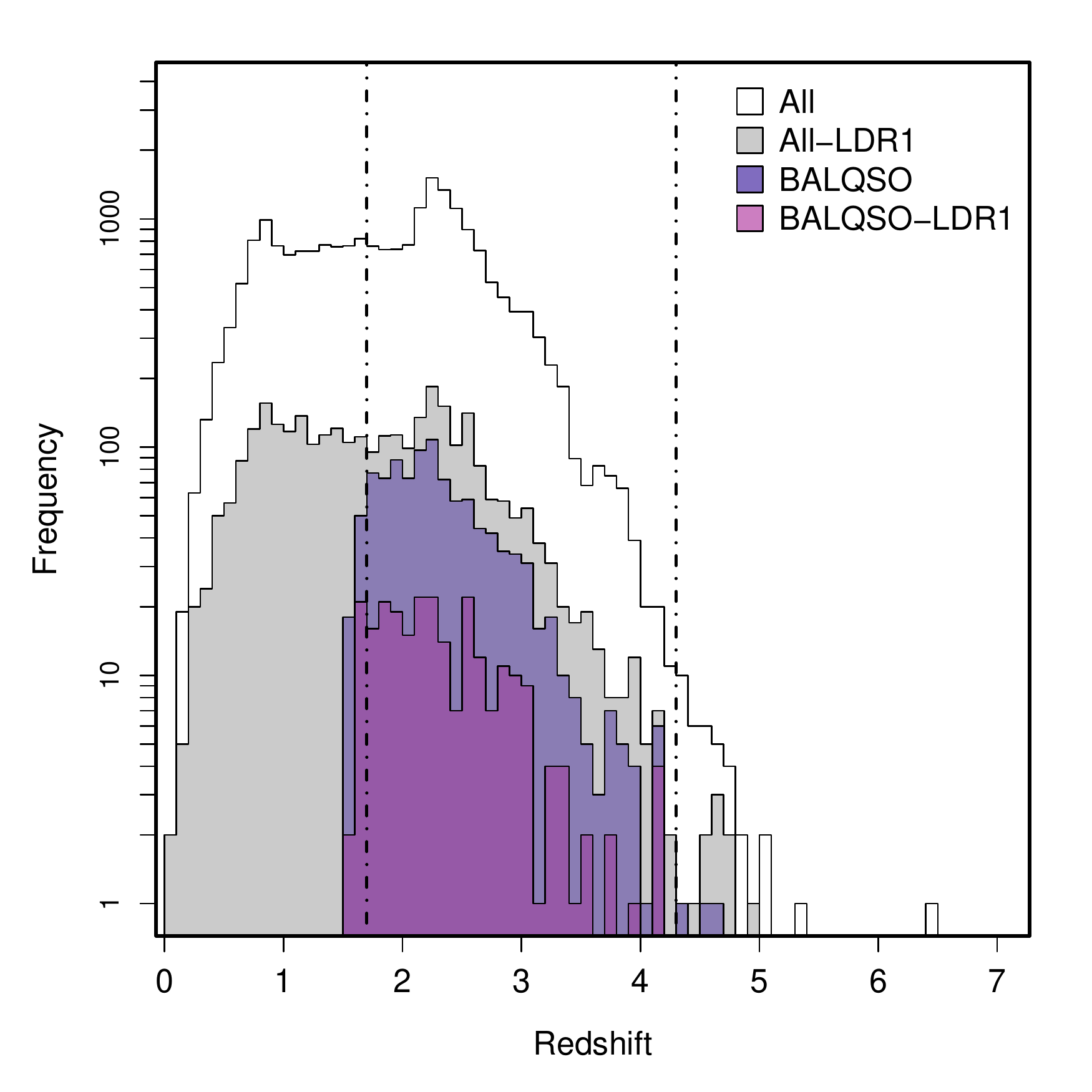}
\includegraphics[width=0.49\textwidth]{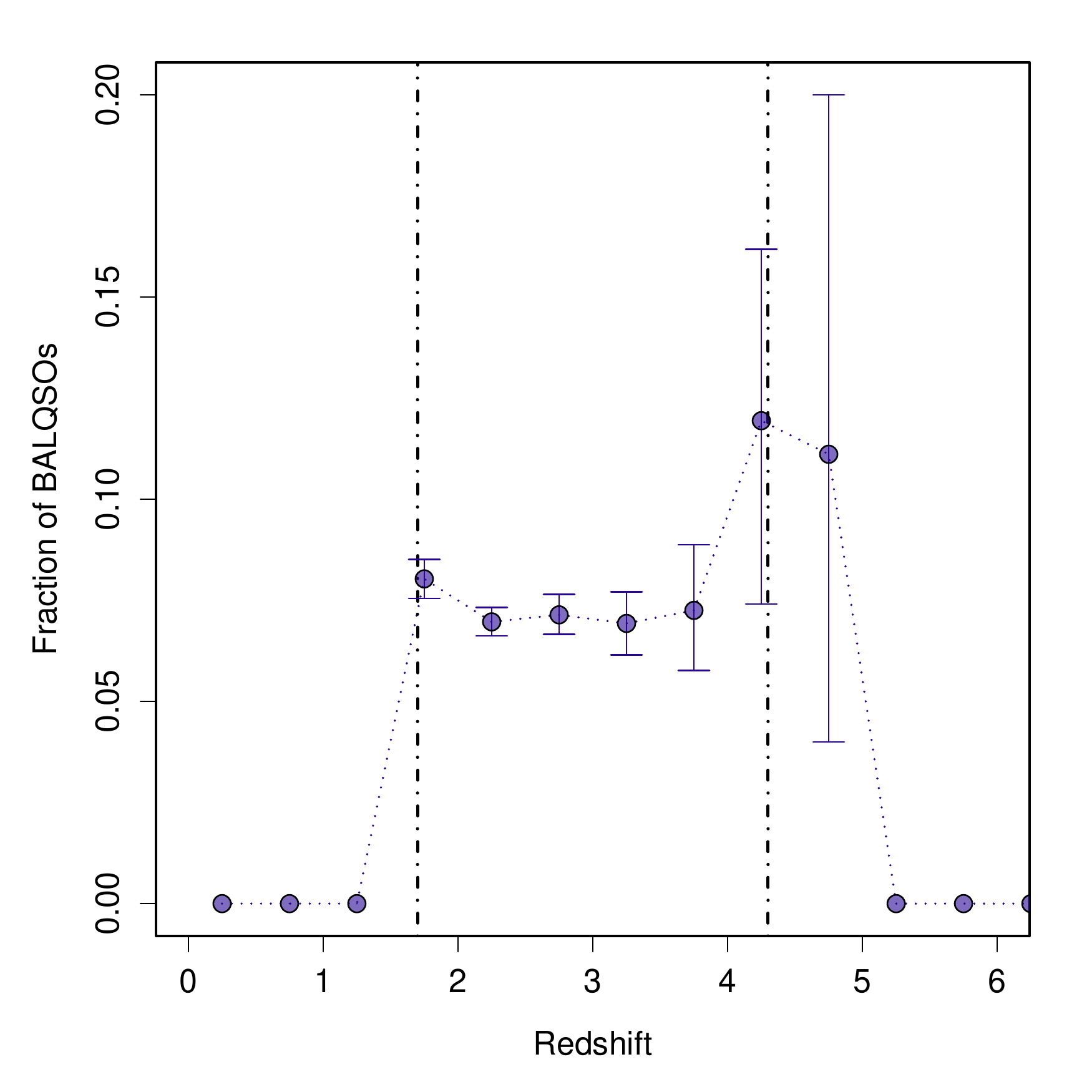}
\caption{\label{fig:f1} {\em Left:} Redshift distributions of the different samples. The vertical dot-dashed lines show the redshift limits for the main sample, where the wavelength coverage is appropriate to measure BI for \civ\ troughs. {\em Right:} Fraction of quasars which are identified as BALQSOs as a function of redshift. The dot-dashed lines are the same as in the left panel.}
\end{figure*}

The quasar sample is summarised in Tab.~\ref{tab:t1}. Although we list the total number of quasars and BALQSOs within the LDR1 MOC, the \textit{fractions} of BALQSOs are not representative since they cannot be identified outside of \zrange . The bottom half of Tab.~\ref{tab:t1} is the final sample that is used in the rest of this study. Eight percent of quasars are identified as BALQSOs, and 12 percent of BALQSOs are LoBALs (one percent of quasars are LoBALs).

\begin{table}
\caption{Quasar sample selection. }
\begin{tabular}{lrl}
\multicolumn{3}{c}{\textbf{All redshifts}} \\ \hline 
 & Number & Fraction \\ \hline 
All quasars & 21812 & 1 \\ 
All quasars, LDR1 detected & 3079 & 0.14 \\ 
BALQSOs & 1045 & 0.048 \\ 
BALQSOs, LDR1 detected & 249 & 0.011 \\ \hline \hline 
\multicolumn{3}{c}{$\mathbf{1.7\leq z \leq 4.3}$} \\ \hline 
All quasars & 12667 & 1 \\ 
All quasars, LDR1 detected & 1615 & 0.13 \\ 
BALQSOs & 976 & 0.077 \\ 
BALQSOs, LDR1 detected & 226 & 0.018 \\ 
LoBALs & 121 & 0.0096 \\ 
LoBALs, LDR1 detected & 48 & 0.0038 \\ \hline 
\end{tabular}
\label{tab:t1}
\tablefoot{The numbers reflect only the quasars which are cross-matched between Pan-STARRS and DR7/DR12, which are over 99 percent of the sources within the LDR1 Multi-Order Coverage map (MOC).}
\end{table}

\subsection{Revisiting 1.4 GHz radio detections}
The FIRST survey also covers the LDR1 area, to a depth of $\sim 0.15\,$mJy at 1400 MHz. Cross-matches with DR7 quasar catalogue are reported in \cite{shen_catalog_2011}. The radio detection fraction for the BALQSOs in DR7 has been well studied \citep[e.g.,][]{stocke_radio_1992,becker_properties_2000,shankar_dependence_2008,dai_intrinsic_2012} and is generally about 20 percent. Within the LDR1 area and \zrange\ there are 119 FIRST matches in DR7, 16 of which are BALQSOs. These matches were found by previous authors using a simple two-step automated radius cross-matching of FIRST sources to the SDSS quasar positions,  using either a 30\sarc\ $\,$radius to find extended radio counterparts, or a $\lesssim$5\sarc\ $\,$radius to find core-dominated radio counterparts.

LoTSS has distinct advantages over this type of radius cross-matching. One major gain is in sensitivity; for a radio source with a typical synchrotron spectral index $\alpha\approx -0.7$, LoTSS is almost an order of magnitude times more sensitive than FIRST. This increased sensitivity not only increases the astrometric accuracy, but also makes it easier to identify the host galaxies of extended radio sources, as the diffuse extended radio emission is often offset from its host galaxy, making an association difficult. Another major gain is the more sophisticated cross-matching of LoTSS sources to PS1 optical counterparts; see Sect.~\ref{sec:ldr1} and \cite{williams_lofar_2018} for more details. 

LoTSS and FIRST have comparable resolutions ($\sim 6$\sarc\ ), and we cross-matched the two radio catalogues with a nearest neighbour algorithm with a radius search of 3\sarc , using only FIRST sources which are not likely to be the sidelobe of a nearby bright source (p\_S$<$0.5). From this radio-radio association, which should not suffer from physical offsets, we used the LDR1 optical counterparts to perform a positional cross-match between Pan-STARRS and SDSS with a search radius of 1\sarc , providing a secure way to associate the FIRST radio sources with SDSS. Doing so yields 381 sources within our sample that are detected in FIRST, 31 of which are BALQSOs. The total number of sources with FIRST detections is thus increased by a factor of 3.2, while the number of BALQSOs doubles. We checked that the values present in DR7 match those of the LOFAR/FIRST matches from this study, and found that all 119 FIRST sources have been identified with the same DR7 quasar.

\section{Results}
\label{sec:results}

\subsection{LDR1 detection fraction and selection effects}
\label{subsec1}
The low-frequency radio detection fraction in our redshift-limited sample can be calculated directly from the values in Tab.~\ref{tab:t1}. Twelve percent of all quasars have LoTSS DR1 counterparts. About 23 percent of BALQSOs have LoTSS DR1 counterparts, while 40 percent of LoBALs have LoTSS DR1 counterparts. The low-frequency radio detection fraction of BALQSOs is therefore about twice that of all quasars (including BALQSOs), and the low-frequency radio detection fraction of LoBALs is even higher. 

We first explore how the LDR1 detection fraction depends on other parameters. Figure~\ref{fig:f2} shows the LDR1 detection fraction as a function of redshift, for the total sample, the subset which are HiBALs, and the subset which are LoBALs. Within the redshift range \zrange\ the mean LDR1 detection fraction for non-BAL quasars is 0.13, while for the BALQSO sub-population, the mean LDR1 detection fraction for BALQSOs is 0.23. The detection rate for BALQSOs is therefore almost twice as high as that for non-BAL quasars. 

\begin{figure*}
\includegraphics[width=\textwidth]{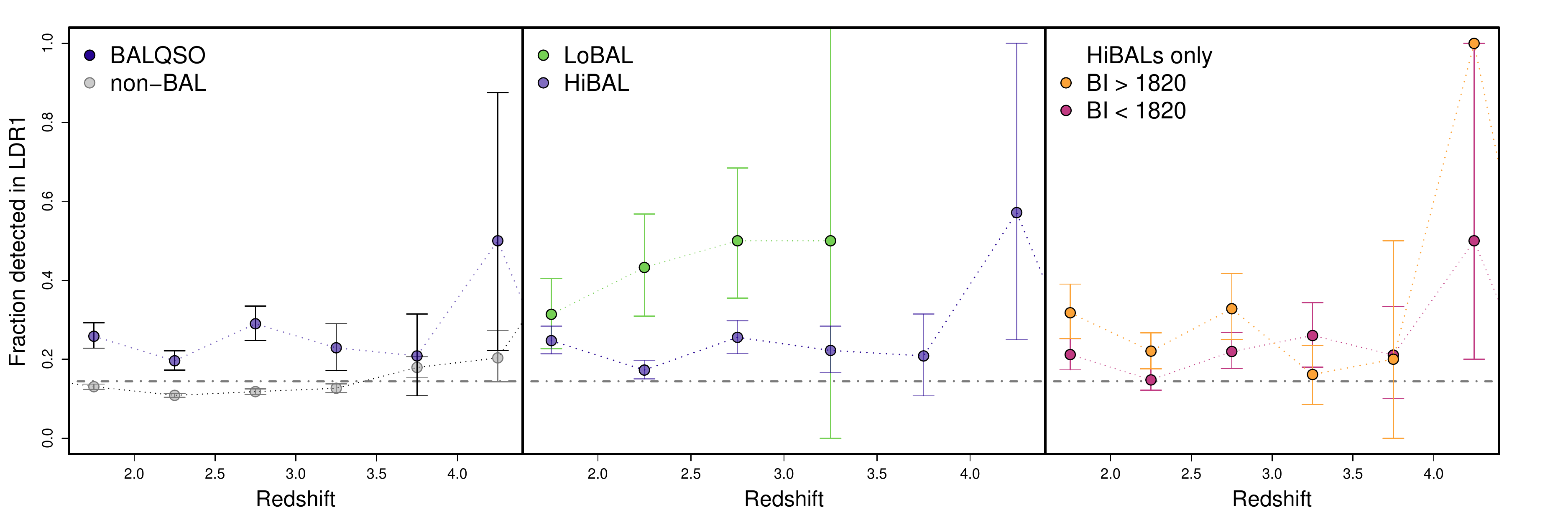}
\caption{\label{fig:f2} Fraction of sources detected in LoTSS DR1 as a function of redshift. The horizontal dot-dashed grey line is the mean LDR1 detection fraction for non-BAL quasars. }
\end{figure*}

Dividing the BALQSOs into LoBALs and HiBALs, and not considering any non-BAL quasars when calculating the detection fractions, we found that the radio detection fraction is 0.40 for LoBALs, and 0.21 for HiBALs. To determine if this is a result simply of the fact that LoBALs tend to have higher BI values, we divide the HiBALs into low and high BI bins, such that the cut-off between the BI bins yields a high-BI bin for HiBALs where the median BI value matches the median BI value of LoBALs. For $z\lesssim3$, the HiBALs with high BI values tend to have higher radio detection fractions, although both low and high BI samples agree within the uncertainties. The radio detection fractions above $z\approx 3$ have much larger uncertainties, and although it appears there is a reversal of the trend, this is not secure. Although it is tentative because of the large uncertainties, it does not appear that large values of BI alone are enough to cause the large radio detection fraction of LoBALs. Larger LoTSS samples in the future will allow us to draw stronger conclusions.

There is clear evidence that the intrinsic fraction of BALQSOs is higher than the observed fraction in optical surveys\citep[e.g.,][]{knigge_intrinsic_2008,dai_2mass_2008,dai_intrinsic_2012}, implying that samples of optically-selected BALQSOs are biased. We therefore investigate whether there are significant selection effects that would bias the radio detection fractions.

There are two main factors that could cause biases in the observed fraction of BALQSOs. First, if the BAL troughs are extreme, the overall flux in a particular optical band used for sample selection can reduce the signal-to-noise ratio and cause BALQSOs to be excluded from a sample. Second, BALQSOs tend to be highly reddened, which can make them more optically faint and less likely to have the high signal-to-noise ratio spectra needed for identification. LoBALs in particular are more reddened than HiBALs, perhaps due to larger quantities of dust near the central AGN \citep{sprayberry_extinction_1992}. 

We first examine the distributions of bolometric\footnote{Bolometric luminosities are from \cite{shen_catalog_2011} and are calculated from monochromatic luminosities using the bolometric corrections from \cite{richards_spectral_2006}. Different monochromatic luminosities are used for different redshift ranges; see \cite{shen_catalog_2011} for details.} and radio luminosities, to assess whether there is a selection effect due to the flux limits of SDSS and LDR1. Figure~\ref{fig:f3} shows the radio luminosity at 144$\,$MHz plotted against the bolometric luminosity, as well as the associated distributions of these parameters for BALQSOs and quasars. The spread in low-frequency radio luminosity remains roughly constant across the entire range of bolometric luminosities, implying that there is no bias towards certain radio luminosities with bolometric luminosity in this sample. A 2D Kolmogorov-Smirnov test cannot rule out the null hypothesis that the underlying distributions of the two populations are the same ($p=0.21$ for non-BAL vs. BALQSO, $p=0.37$ for LoBAL vs. HiBAL). It is worth noting that selecting only LoBALs from the BALQSO sample does not significantly shift the bolometric luminosity distribution to the right, indicating that we are not missing a significant fraction of fainter LoBALs. 

\begin{figure}
\includegraphics[width=0.5\textwidth]{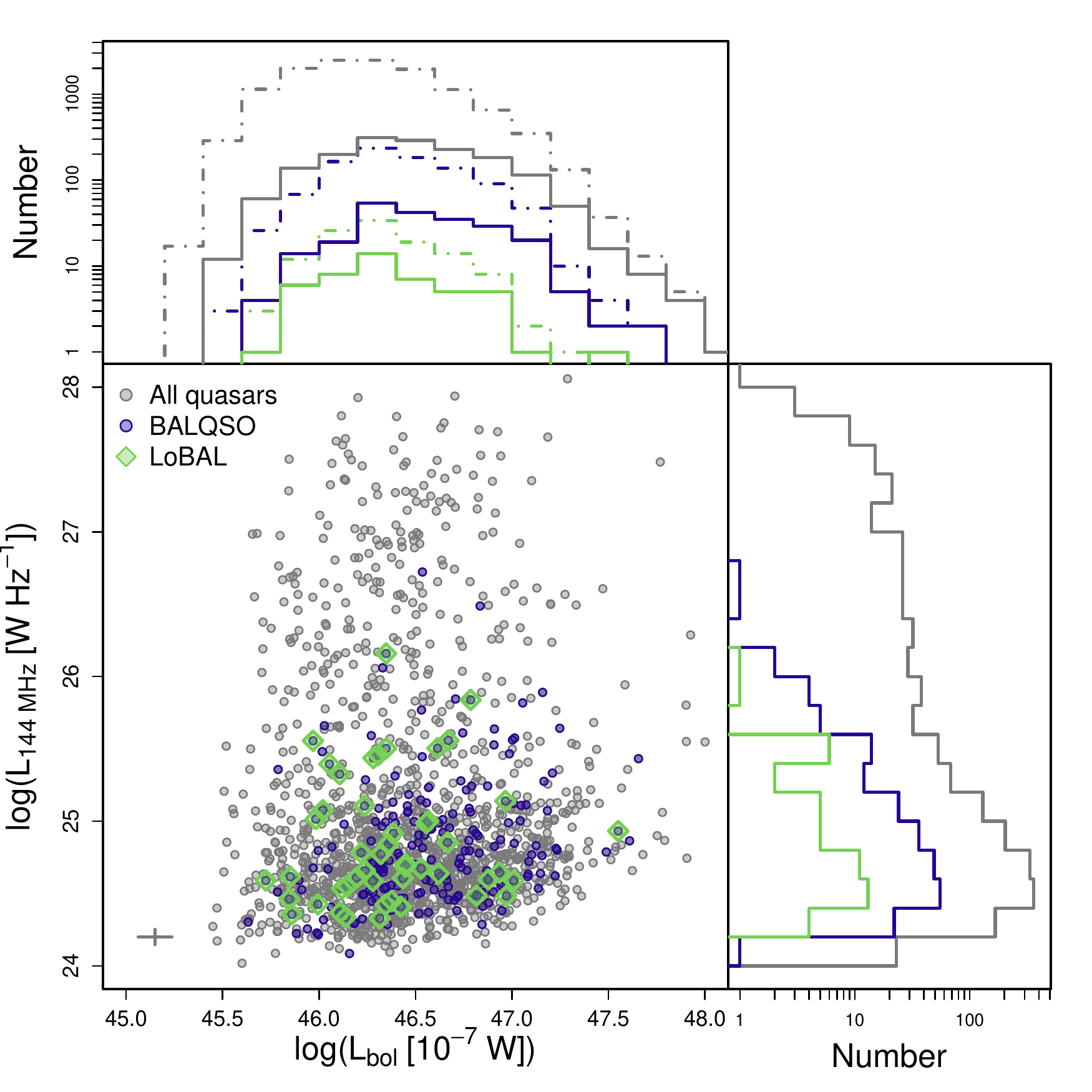}
\caption{\label{fig:f3} Radio luminosity vs. bolometric luminosity. Light gray points are all quasars with LoTSS detections, and purple points are those classified as BALQSO. The median errors are shown by the cross in the bottom left of the plot. The top panel shows the distributions of bolometric luminosity: solid lines represent the LoTSS-detected samples, while dot-dashed lines represent all sources regardless of LoTSS-detection. The right panel shows the distributions of radio luminosity for LoTSS-detected sources. The line colours are the same as in the scatter plot.}
\end{figure}

We address radio-loudness in a later sub-section, but for now we note that there is no strong correlation between low-frequency radio luminosity and optical bolometric luminosity; i.e., the spread in radio luminosity remains similar from the low to high bolometric luminosity end of Figure~\ref{fig:f3}. The lack of a strong correlation between radio luminosity and bolometric luminosity suggests that the radio properties of optically bright and faint BALQSOs are similar. This is in agreement with \cite{morabito_unveiling_2014}, who compared optically faint and optically bright BALQSOs and showed that there is no strong evidence for difference in absorption line properties. As radio emission is unaffected by dust, the presence or absence of dust will not impact whether we detect radio emission if it is present in reddened BALQSOs.

A small fraction of quasars were selected for inclusion in the SDSS sample from their compact morphology in FIRST. As we are interested in the radio detection fraction, this could potentially bias our results. We removed these objects using the targeting flags in DR12 and verified that this does not change our results. We did find that LoBALs were slightly preferentially selected by FIRST detections (at the $\sim$10 percent level), perhaps because the BAL features affect the optical colours, moving some LoBALS out of the SDSS colour-selection space, but again our results do not change within the uncertainties reported in this paper. 

To further test if selection effects are important, we plot in Figure~\ref{fig:f4} the ratio of the radio detection fractions of BALQSOs and LoBALs to non-BAL quasars as a function of bolometric luminosity. The ratio of LoTSS detections decreases rather than increases with increasing luminosity, indicating the increased radio detection fraction in BALQSOs and LoBALs cannot be explained by a selection effect which favours non-BAL quasars over BALQSOs at low bolometric luminosities. Due to the smaller number of LoBALs in the sample, the uncertainties are larger than for all BALQSOs, but the trend is similar. An even steeper trend might be possible, with a higher LoBAL to non-BAL radio detection fraction at lower bolometric luminosities, but a larger sample is needed to confirm this. 

\begin{figure}
\includegraphics[width=0.5\textwidth]{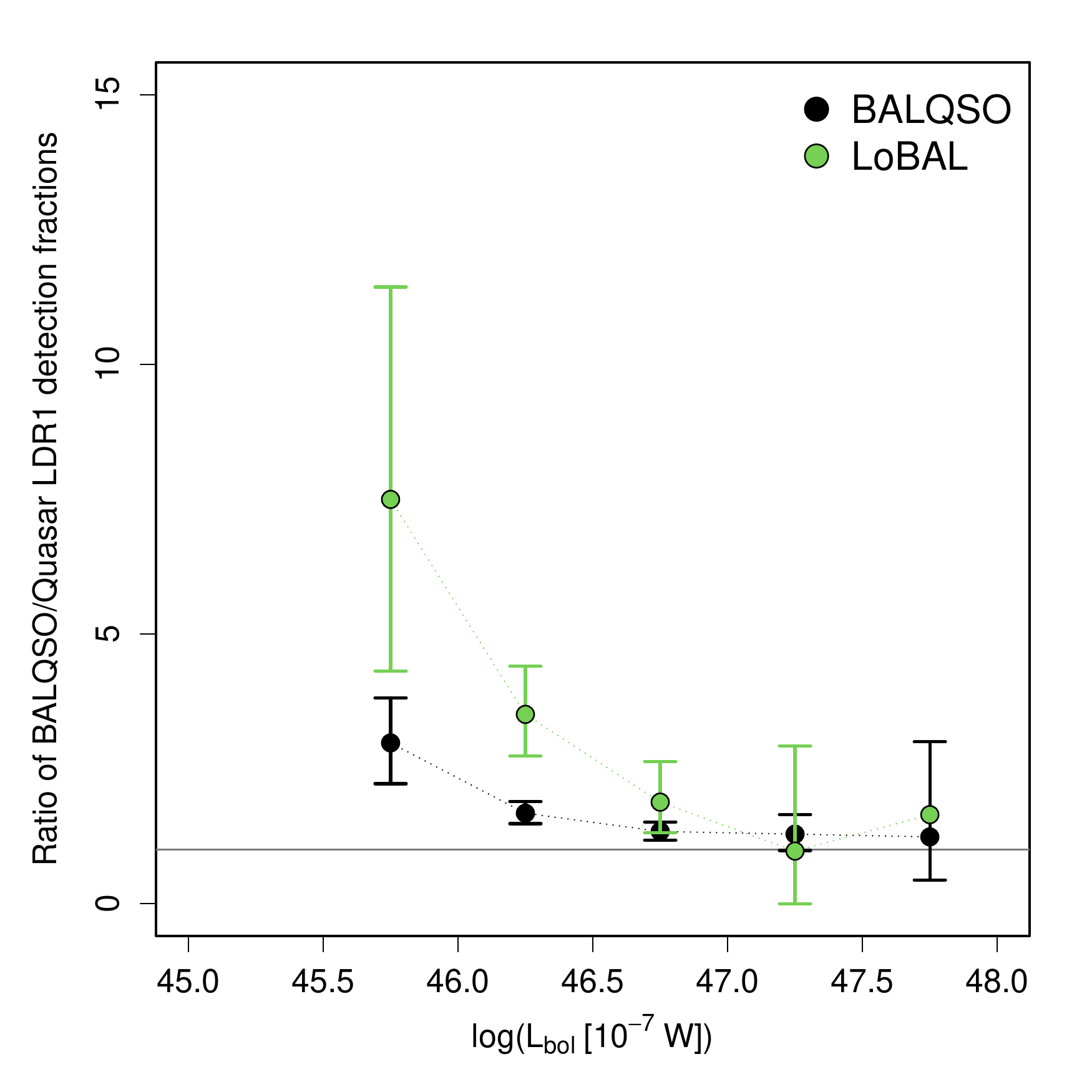}
\caption{\label{fig:f4} Ratio between the LDR1 detection fraction of BALQSOs to quasars, as a function of bolometric luminosity. }
\end{figure}

\subsection{Fraction of BALQSOs as a function of radio luminosity}
\label{subsec2}
For the subset of quasars that are detected in LDR1, we consider the BALQSO fraction as a function of low-frequency radio luminosity (assuming $\alpha=-0.7$), see Figure~\ref{fig:f5}. Within the error bars, there is no increase of the fraction of quasars which are BALQSOs in all except the highest radio luminosity bin. The final bin is inconsistent with the rest of the data; this may be due to the fact that LDR1 is limited in sky coverage and there are only three BALQSOs in this bin, although the uncertainties on the {\em fraction} remain small since the number of non-BAL quasars does not decline as rapidly with increasing radio luminosity. As LoTSS observes more of the sky, it will be interesting to revisit these results.

\begin{figure}
\includegraphics[width=0.5\textwidth]{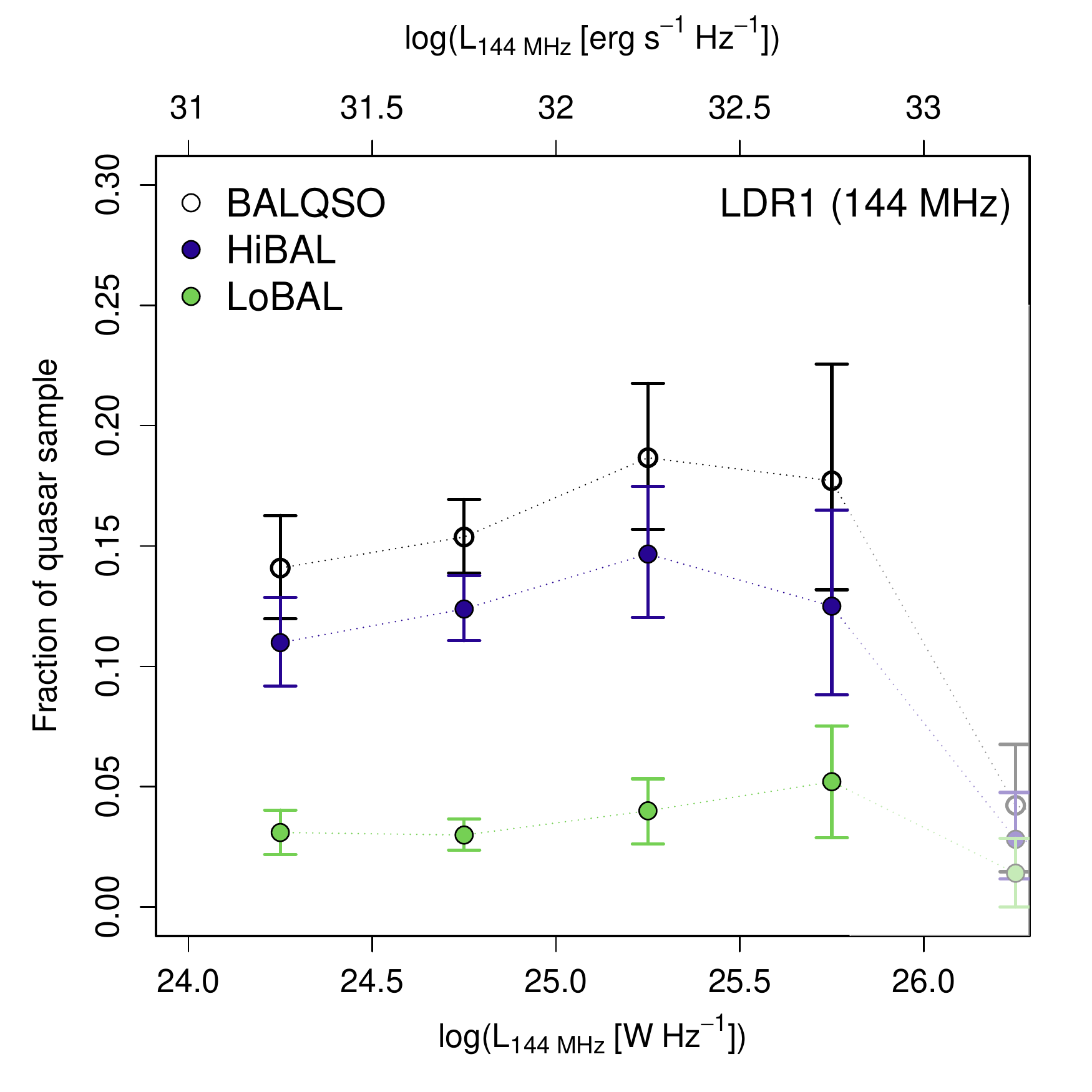}
\caption{\label{fig:f5} Fraction of BALQSOs and LoBALs in the overall quasar sample as a function of radio luminosity at 144$\,$MHz. The abscissa are luminosity bin midpoints. As discussed at the beginning of Sect.~\ref{subsec2}, the highest luminosity bin may be heavily biased by the small number of sources and the limited sky coverage; these points have been shaded out.}
\end{figure}

The trend at 144$\,$MHz for the BALQSO fraction of radio-detected quasars to remain constant with radio luminosity is inconsistent with what has been seen with FIRST detections \citep[e.g.,][]{shankar_dependence_2008}. To check this, we made the same plot as Fig~\ref{fig:f5} but using FIRST detections rather than LDR1 detections. This is shown in the left panel of Figure~\ref{fig:f6}. Even using our expanded sample of FIRST detections rather than the DR7-FIRST associations, there are only 381 FIRST sources rather than the 1$\,$582 LDR1 sources. We find, consistent with previous results, that the BALQSO fraction decreases with increasing radio luminosity. Our values are consistent within the uncertainties with \cite{shankar_dependence_2008}, although our bin sizes and uncertainties are larger. To check that the constant BALQSO fraction with low-frequency radio luminosity for LoTSS-detected sources is not driven by the much larger sample size, we repeated Figure~\ref{fig:f5} for LoTSS-detected sources but only used sources which had FIRST detections. This is shown in the right panel of Figure~\ref{fig:f6}, and although the uncertainties are large, we find again that the BALQSO fraction is constant within the uncertainties, again with the exception of the highest luminosity bin, which is likely to be limited by the relatively small sky coverage of LDR1. 

\begin{figure*}
\includegraphics[width=0.49\textwidth]{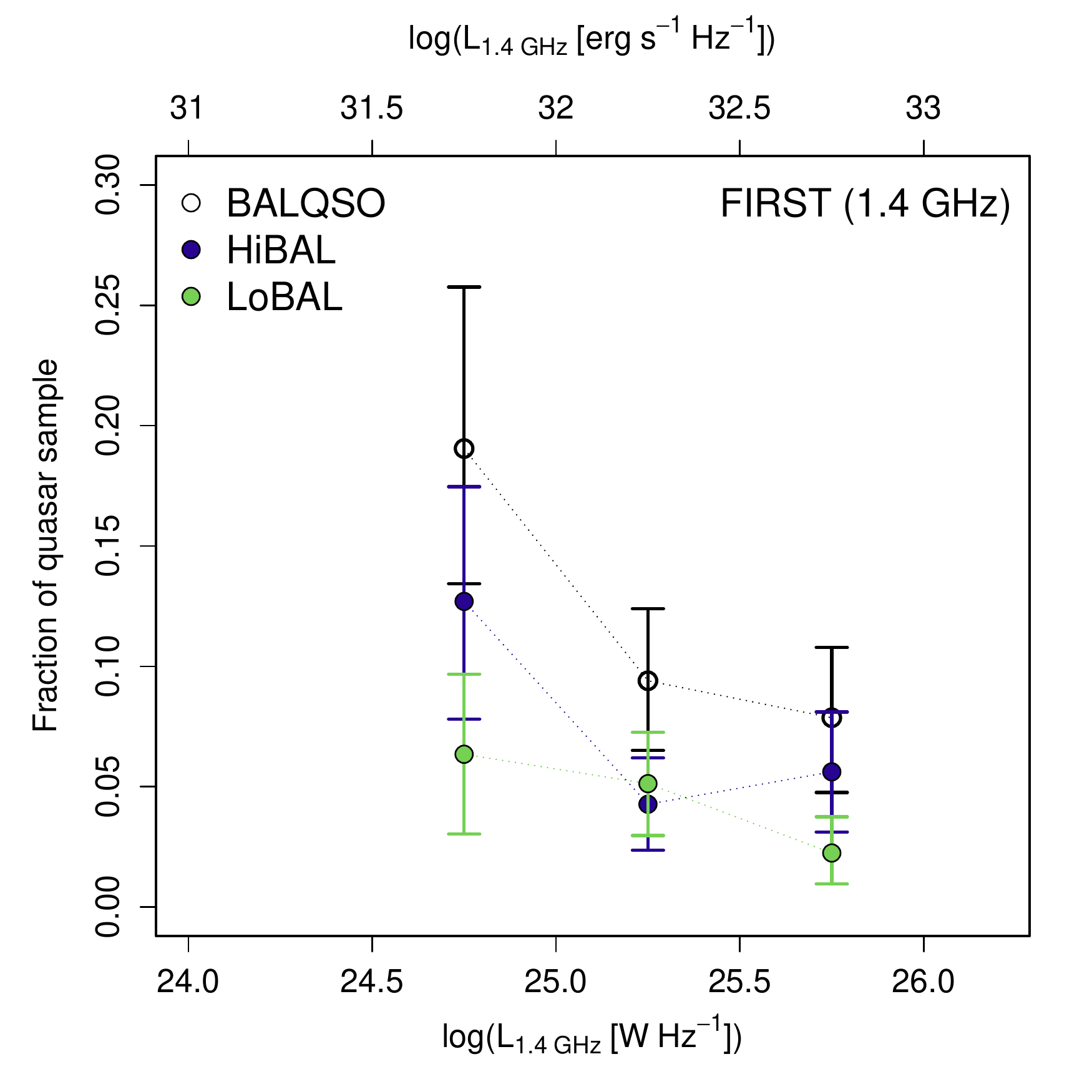}
\includegraphics[width=0.49\textwidth]{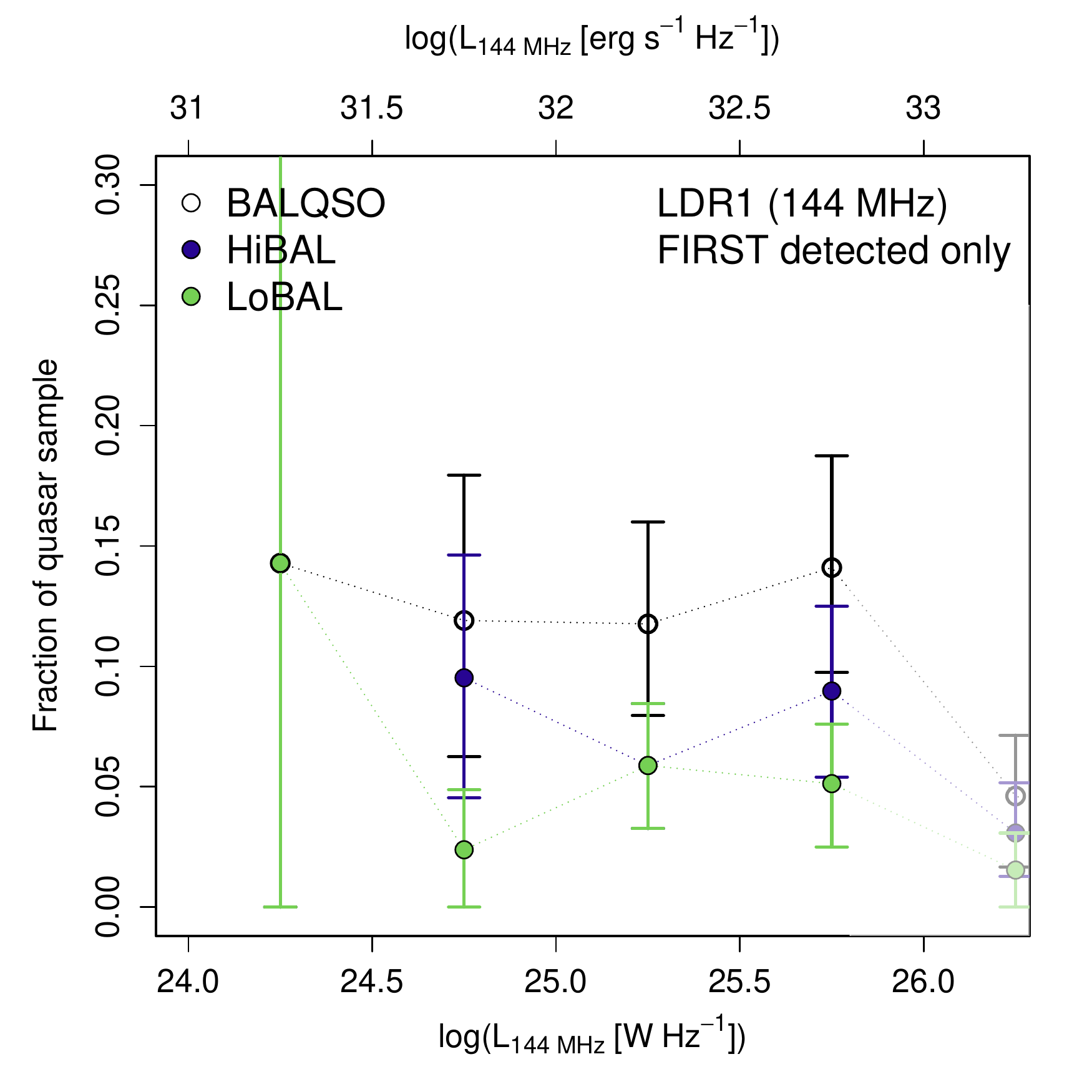}
\caption{\label{fig:f6} {\em Left:} Fraction of BALQSOs, LoBALs, and HiBALs in the overall quasar sample as a function of radio luminosity at 1.4 GHz for the sources with FIRST counterparts. {\em Right:} Fraction of BALQSOs, LoBALs, and HiBALs in the overall quasar sample as a function of radio luminosity at 144$\,$MHz, but for only the sources with FIRST counterparts. As discussed at the beginning of Sect.~\ref{subsec2}, the highest luminosity bin may be heavily biased by the small number of sources and the limited sky coverage; these points have been shaded out. }
\end{figure*}

The BALQSO/LoBAL/HiBAL fractions of radio-detected quasars decrease with increasing radio luminosity at 1.4$\,$GHz and are constant with increasing radio luminosity at 144$\,$MHz, even when using only the LoTSS-detected sources which have FIRST counterparts. This implies that the two observing frequencies may be tracing different sources of radio emission. However, we do note that small-number statistics may still be dominant, as the first power bin of 10$^{24-25}$\wphz\ only has a single (LoBAL) BALQSO in it.

\subsection{Radio Loudness}
\label{subsec3}
Radio loudness is generally defined by the ratio of radio to optical luminosity \citep[e.g.,][]{kellermann_vla_1989}, $R = L_{\textrm{radio}} / L_{\textrm{optical}}$. Typically this is done using the radio luminosity at 5$\,$GHz and the optical luminosity at B band ($\sim 450\,$nm). Ratios of $R>10$ are generally considered to be radio loud. Here we use $L_{\textrm{144 MHz}}$ for the radio luminosity and the luminosity derived from the PS1 $g$-band ($\lambda_{\textrm{eff}}=481\,$nm), $L_{g}$ for the optical luminosity. We do not convert the radio luminosity to 5 GHz because we do not have enough information on the radio spectral index values (spectral index values are discussed in Sect.~\ref{subsec5}) to know that this can be done correctly. Using a typical spectral index of $\alpha=-0.7$ to extrapolate the radio-loud cut-off from 5$\,$GHz to 144$\,$MHz, we define $\log(R_{\textrm{144 MHz}})>2$ as radio-loud. 

\begin{figure}
\includegraphics[width=0.5\textwidth]{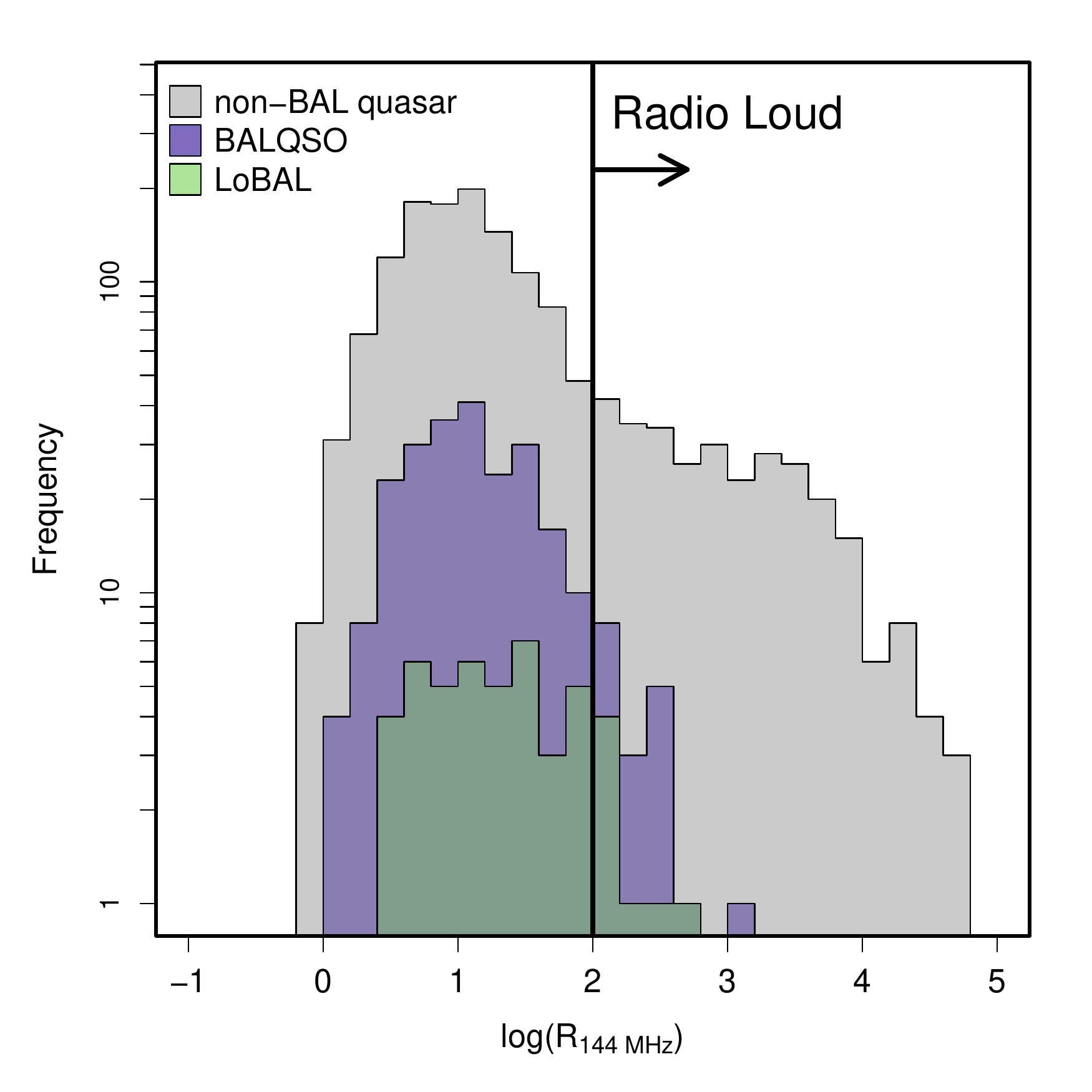}
\caption{\label{fig:f7} Distributions of radio loudness of non-BAL quasars, BALQSOs, and LoBALs.}
\end{figure}

The distributions of \logr\ are shown in Figure~\ref{fig:f7}. It is clear that the distribution of radio loudness for non-BAL quasars extends to higher values of \logr , while BALQSOs and LoBALs all have values of \logr $\,<2.5$. In the non-BAL quasar distribution, there is slight evidence for two peaks, one at $\sim 1$ and one at $\sim 3.5$. This could be evidence for two different populations of non-BAL quasars: one which has radio to optical luminosity ratios similar to BALQSOs, and one in which the radio luminosity is stronger. Using the definition of \logr $\,>2$ we find that 7 LoBALs and 11 HiBALs are radio-loud, for a total of 18 radio-loud BALQSOs. The majority of BALQSOs would be classified as radio-quiet, although the distribution of \logr\ smoothly extends across the border of radio-quiet/radio-loud and is not a clear dichotomy.

We find, similar to \cite{becker_first_2001}, that there is a smooth distribution across the historically-defined division between radio-quiet and radio loud, with a sharp drop in the BALQSO distribution above a certain radio-loudness value.  \cite{becker_first_2001} found this to be at \logrstar $>2$, while we see the a drop in BALQSOs around \logr $\sim 2.5$ and LoBALs just above \logr $\sim 2$. This drop is not seen in the non-BAL quasars. Qualitatively, our results agree with \cite{becker_first_2001}. It is not clear whether this drop off is physically meaningful, or due to the fact that both the population of BALQSOs and the population of extremely powerful radio sources (which are more likely to be radio-loud) {\em individually} have low number density; the {\em combination} of these populations will have an even lower number density. The sharp drop in radio loudness could therefore be due to the limited sky coverage of LDR1 -- which will only become clear as LoTSS covers larger areas of the sky.

\subsection{Radio sizes}
\label{subsec4}
The LDR1 catalogue provides the sizes of radio sources. We use either the LOFAR Galaxy Zoo size, if it exists, or 2 times the full width at half maximum (FWHM) of the deconvolved major axis (for more details on why the factor of 2 is appropriate see Sect.~2.1 of Hardcastle et al., submitted) to determine the projected largest linear size (LLS). FIRST also provides the FWHM of the deconvolved Gaussian fit to the radio source, which we multiply by 2. Figure~\ref{fig:f8} shows the distributions of the logarithm of the largest linear sizes for both LDR1 and FIRST, as well as the LDR1/FIRST LLS ratio. This was divided into non-BAL quasars and BALQSOs for comparison of the two populations, and further divided into resolved and unresolved for LDR1. It is clear that for both populations, the median LLS is larger for LDR1. The BALQSO radio sizes also tend to be smaller than the non-BAL quasar radio sizes. The size ratios for both populations are similar, with the exception of a long tail towards higher size ratios for the non-BAL quasars. A Kolmogorov-Smirnov test cannot rule out the null hypothesis that the underlying distributions of size ratios are the same ($p$-value$\,=0.15$).

The majority of sources are unresolved in LDR1, and only seven BALQSOs, one of which is a LoBAL, are resolved, although still single-component radio sources. We show the resolved BALQSOs as well as a selection of resolved non-BAL quasars, and unresolved BALQSOs and non-BAL quasars in Appendix~\ref{appendix1}. The fact that all but seven BALQSOs are unresolved at 6\sarc\ \ is consistent with previous results \citep[e.g.,][]{becker_properties_2000,dipompeo_survey_2011} which find that the radio emission from BALQSOs is compact. While we cannot determine the morphology of the radio emission at the resolutions in LDR1 and/or FIRST, the fact that the LDR1 radio sizes are on average larger than the FIRST radio sizes is suggestive of the radio emission being due to synchrotron-dominated jets with an AGN core \citep[e.g.,][]{ceglowski_vlbi_2015,bruni_parsec-scale_2013,liu_compact_2008}. LOFAR is capable of sub-arcsecond resolution \citep[e.g.,][]{varenius_subarcsecond_2015,morabito_lofar_2016}, and follow-up studies to resolve the morphology of the low-frequency radio emission will be informative. 

\begin{figure*}
\includegraphics[width=\textwidth]{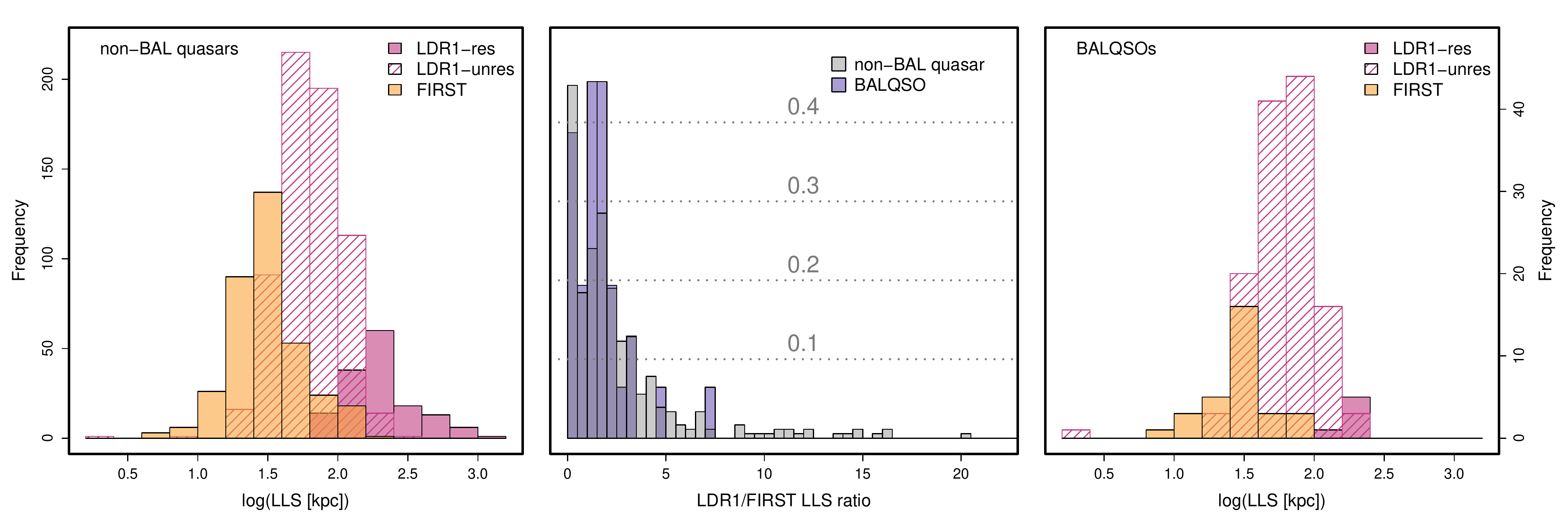}
\caption{\label{fig:f8} {\em Left:} Distribution of largest linear sizes in LDR1 (pink) and FIRST (orange) for non-BAL quasars. The pink hatched area represents unresolved sources in LDR1, and the solid pink area represents resolved sources in LDR1  {\em Centre:} Normalised distributions of LDR1/FIRST LLS ratios for non-BAL quasars (grey) and BALQSOs (purple). There are five sources with extreme ratios ($25<$LLS$<70$) which are not pictured here. {\em Right:} Distribution of largest linear sizes of BALQSOs in LDR1 (pink) and FIRST (orange). The hatched and solid areas are for unresolved and resolved sources in LDR1, respectively.}
\end{figure*}

The observed characteristics of radio jets can be linked to either evolution or orientation, although orientation effects will always be present. Radio jets have historically been used as a proxy for orientation \citep[e.g.,][]{barthel_is_1989,morabito_investigating_2017}, as their projected linear sizes will depend on the jet angle to the line of sight. If a class of objects has jets of approximately the same size (or size distribution), those oriented with the jets in the plane of the sky will have larger projected LLS size ratios, and those with jets oriented along the line of sight will have smaller projected LLS size ratios. If balnicity is also a proxy for orientation, there should be a correlation between BI and the LLS size ratios. We plot this in Figure~\ref{fig:f9} for HiBALs and LoBALs. While both the LoBAL and HiBAL samples show a suggestion of an anticorrelation between BI and LDR1/FIRST LLS size ratio (stronger for LoBALs than HiBALs), there is not enough data for this to be significant. As LoTSS expands to cover more of the sky so will our sample size, and this will be an interesting question to revisit. In particular, if we find that a full range of jet orientations is possible in BALQSOs, this might mean that the BAL winds have a covering fraction of close to 1, implying that they are at a special quasar evolutionary stage. Alternatively, BAL winds could emerge at a range of angles as suggested by, e.g., \cite{yong_properties_2018}.

\begin{figure}
\includegraphics[width=0.5\textwidth]{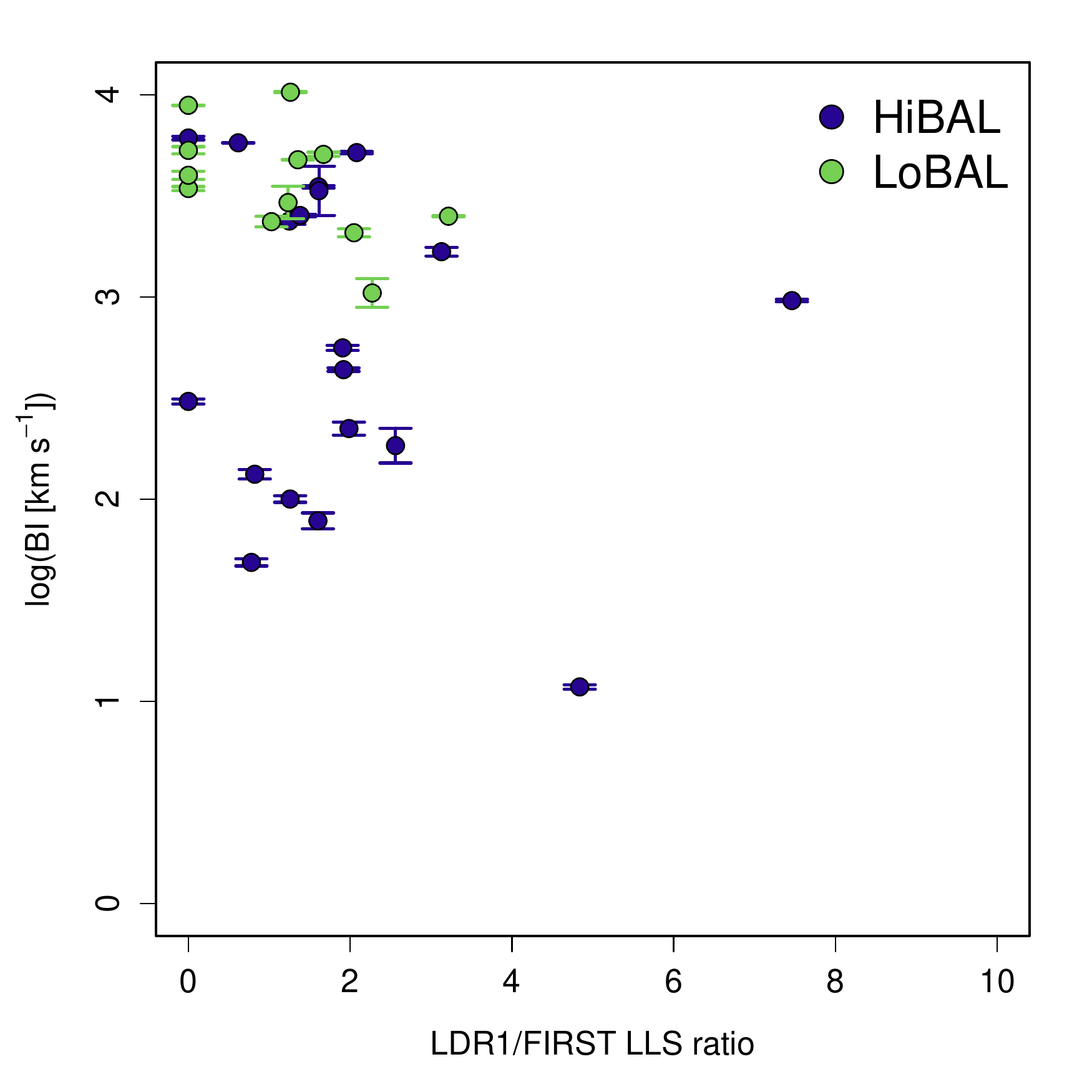}
\caption{\label{fig:f9} Balnicity index (BI) vs. LDR1/FIRST LLS ratios for HiBALs (blue) and LoBALs (green).}
\end{figure}

\subsection{Spectral properties}
\label{subsec5}
We next examined the spectral properties, using both LDR1 and FIRST flux density measurements. We caution the reader that LDR1 is approximately 10 times deeper than FIRST, and therefore will be biased towards sources with flatter spectral indices. We tried cross-matching the sample with WENSS to provide an intermediate frequency, but the only matches were for non-BAL quasars. Figure~\ref{fig:f10} shows the LDR1 vs. FIRST measured radio luminosities, and Tab.~\ref{tab:t2} shows the weighted median of the spectral index values (calculated only for sources detected in both LDR1 and FIRST) for different sub-samples, with bootstrapped errors. In Figure~\ref{fig:f10} we have drawn two lines of constant spectral index, at $\alpha=-0.5$ and $\alpha=+0.5$. The majority of sources detected in both surveys lie within these two lines, although there are more sources clustered around $\alpha=-0.5$ than around $\alpha=+0.5$. Upper limits for sources detected in LDR1 but not FIRST are shown as arrows\footnote{The upper limits were calculated by finding the limiting spectral index assuming the FIRST detection threshold of 1$\,$mJy.} in Figure~\ref{fig:f10}, and do not exclude a large portion of the expected parameter space. The median values, listed in Tab.~\ref{tab:t2} show a tendency for BALQSOs to have flatter spectral indices than non-BAL quasars, with HiBALs having spectral indices consistent with flat spectra. Without an intermediate frequency measurement, it is difficult to know whether the flat spectral index values are caused by truly flat spectra, or spectra that peak at intermediate frequencies as BALQSOs are known to do. Follow up observations with, for example, the Giant Metre-wave Radio Telescope at 610 MHz will help determine the intrinsic shape of the radio spectra.

\begin{table}
\caption{Weighted median spectral index values and bootstrapped uncertainties, for sources detected at both frequencies.}
\begin{tabular}{lc}
 Sample & Weighted median$\pm$bootstrap uncertainty \\ \hline 
All quasars & $-0.65\pm0.087$ \\ 
non-BAL quasars & $-0.65\pm0.091$ \\ 
BALQSOs & $-0.094\pm0.25$ \\ 
LoBALs & $-0.26\pm0.14$ \\ 
HiBALs & $-0.094\pm0.32$ \\ 
\end{tabular}
\label{tab:t2}
\end{table}

\begin{figure}
\includegraphics[width=0.5\textwidth]{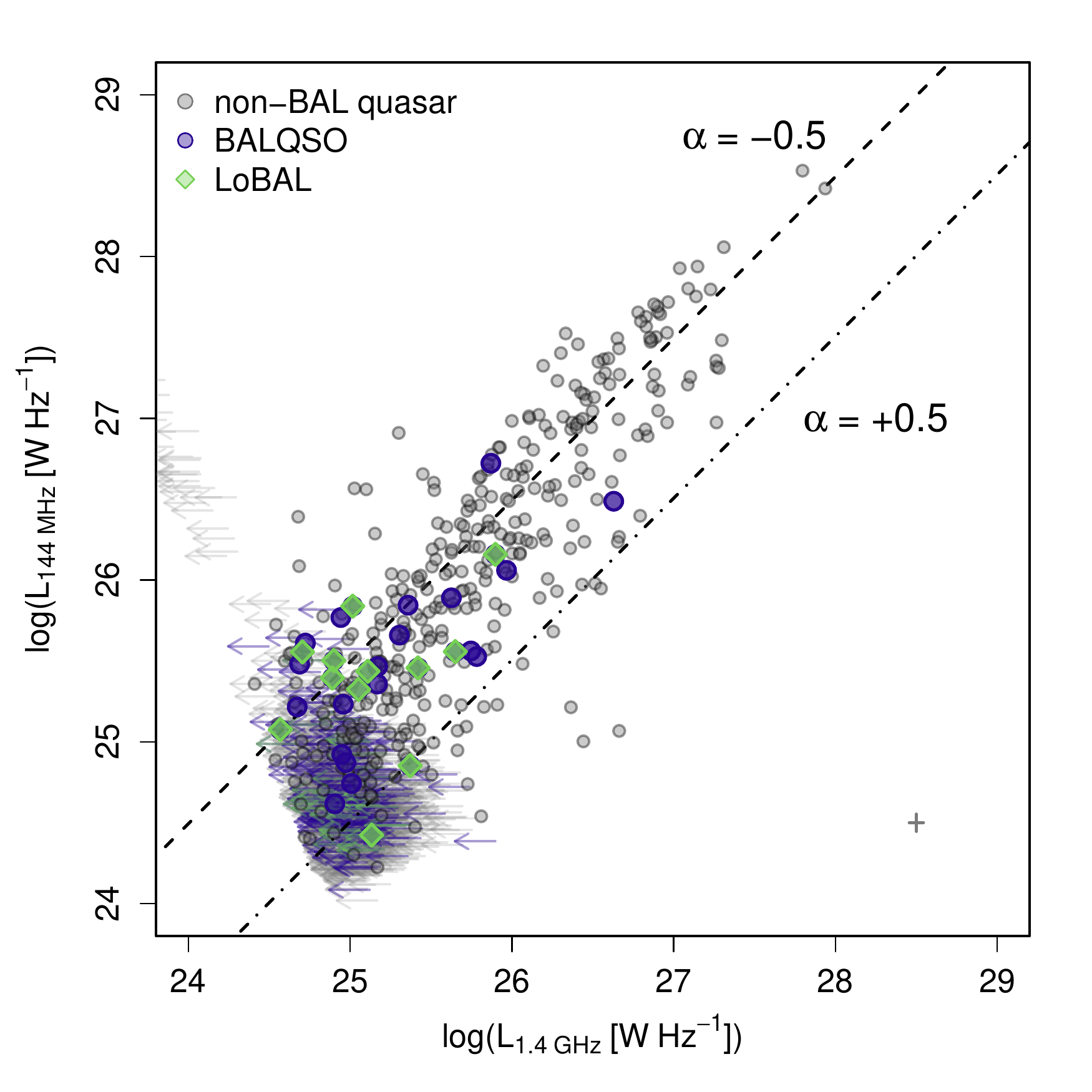}
\caption{\label{fig:f10} Radio luminosity from LDR1 vs. FIRST measurements. Lines of constant spectral index are drawn on the plot, with the appropriate labels. Upper limits for sources are detected in LDR1 but not FIRST are shown as left-pointing arrows. The median uncertainties for detections (i.e., not upper limits) are shown as a cross in the bottom right corner of the plot.}
\end{figure}

\subsection{LoBAL fractions}
\label{subsec6}
Finally, we investigated the fraction of BALQSOs which are LoBALs as a function of radio luminosity. LoBALs are thought to be either normal BALQSOs viewed along a particular line of sight or the progenitors of HiBALs/non-BAL quasars. We plot the fraction of LoBALs as a function of radio luminosity in Figure~\ref{fig:f11} for both LDR1 and FIRST radio luminosities. To aid the comparison, we have shifted the LDR1 radio luminosity abscissa by an amount equivalent to the median spectral index of the sample, although we stress that low- and high-frequency radio luminosities may be related to different radio emission processes, and we should look at general trends only when comparing the fraction of LoBALs. Although the number of FIRST-detected LoBALs in DR7 is small, we wish to compare our results with  previous studies. We therefore show the fraction of LoBALs in the FIRST sample  adjusted by the ratio of LDR1/FIRST cross-matched detections to the historic DR7 FIRST detections. We find that our fractions of LoBALs for the FIRST-adjusted values in Figure~\ref{fig:f11} agree well with Figure 7 in \cite{dai_intrinsic_2012}.

\begin{figure}
\includegraphics[width=0.5\textwidth]{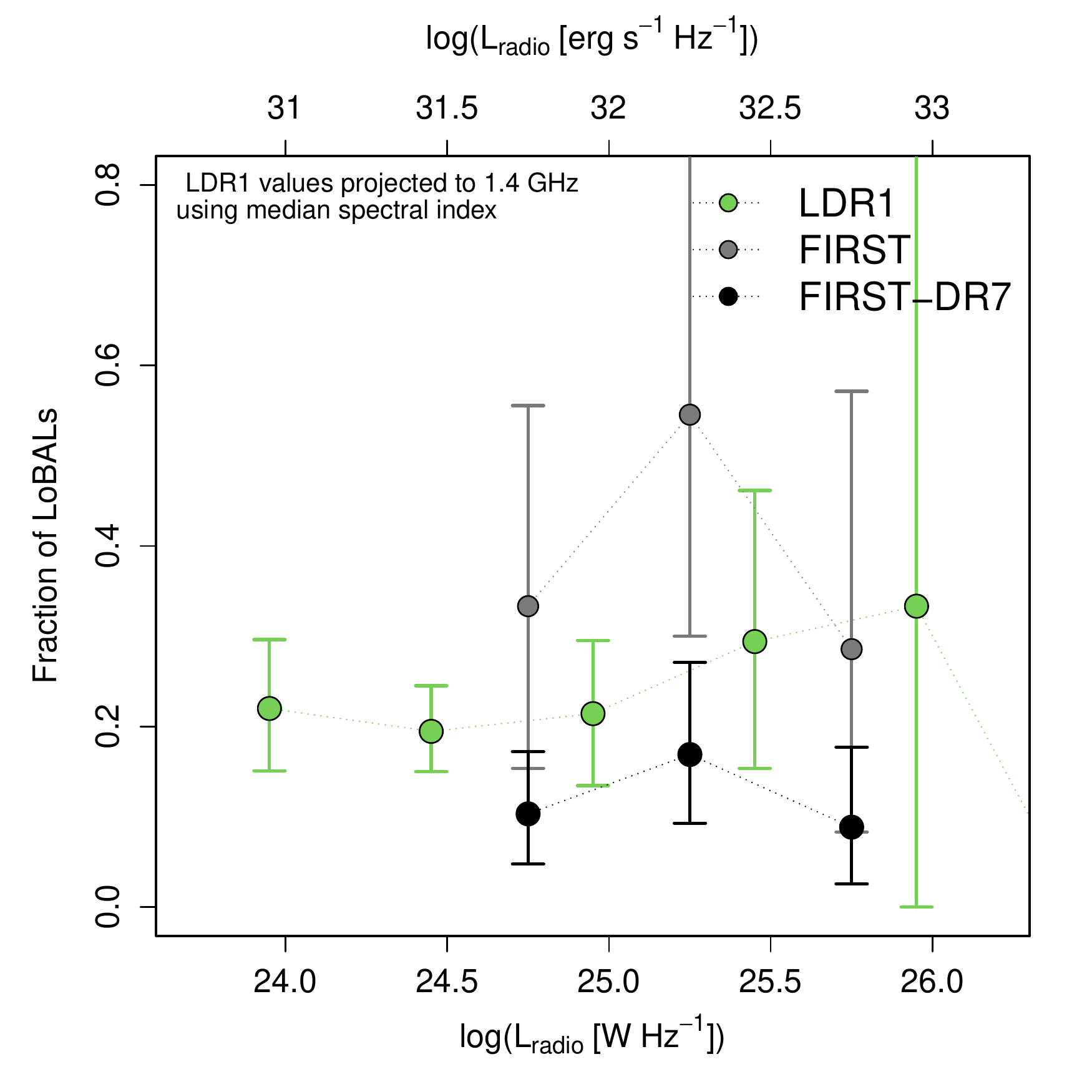}
\caption{\label{fig:f11} Fraction of BALQSOs which are LoBALs, as a function of radio luminosity. We have included fractions as a function of both LoTSS and FIRST powers. To aid the comparison, we have shifted the LoTSS radio luminosity abscissa by an amount equivalent to the median spectral index between the LoTSS/FIRST samples. Both observed and adjusted FIRST values are plotted, where the adjusted values have been reduced by the fraction of DR7 FIRST detections to LoTSS/FIRST cross-matched detections.}
\end{figure}

The LoBAL fraction in the FIRST sample remains constant with radio luminosity, although the uncertainties are large. The median values (and bootstrapped uncertainties) are $0.20\pm 0.01$ for LDR1, $0.33\pm 0.08$ for FIRST, and $0.11 \pm 0.03$ for FIRST-DR7 samples. The two main results from this are: the fraction of LoBALs is lower amongst LoTSS-detected sources than FIRST-detected sources, and that there is a significant number of LoBALs which were missed in previous FIRST samples of BALQSOs. 

\subsection{Absorption line properties}
\label{subsec7}
BALQSOs are identified by their BI, which is defined in terms of the strength of the broad absorption lines. If the radio emission in BALQSOs is related to the same processes that drive the broad absorption lines, we would expect to find correlations between radio properties and BI. In Figure~\ref{fig:f12} we plot the  balnicity index (BI) as a function of radio loudness, and the radio luminosity at both 144$\,$MHz and 1.4$\,$GHz. In a sample of 29 BALQSOs, \cite{becker_properties_2000} found an anti-correlation of BI and $L_{\textrm{1.4 GHz}}$ for HiBALs (a Spearman rank coefficient of -0.85 and probability of 6$\times10^{-5}$), and no correlation for LoBALs. We tested for correlations by calculating Spearman's correlation coefficient and $p$-values. All of the results had low significance, indicating that their BI is not correlated with \logr , $L_{\textrm{144 MHz}}$, or $L_{\textrm{1.4 GHz}}$; see Tab.~\ref{tab:t3} for the results. This suggests that the anti-correlation reported by \cite{becker_properties_2000} was likely due to a combination of the small sample size of the FIRST-detected BALQSOs  and the inclusion of objects with $\mathrm{BI}=0$ in the BALQSO sample. However, we caution that even with the expanded FIRST sample in this paper, which is more than a factor of three larger than the DR7 FIRST sample, the number of BALQSOs with measured BI and radio detections at 1.4$\,$GHz is still only 31 sources; this compares to 22 sources with $\mathrm{BI}>0$ in the \cite{becker_properties_2000} sample. As LoTSS expands to covers larger areas, revisiting the relationship between BI and radio properties will be instructive.

\begin{table}
\caption{Spearman's rank correlation coefficients between HiBAL and LoBAL radio properties and BI.}
\begin{tabular}{lcc}
 & HiBAL & LoBAL \\ \hline 
log($R_{144}$) & -0.022, $p=0.76$ & -0.23, $p=0.11$ \\ 
L$_{\textrm{144 MHz}}$ & -0.076, $p=0.32$ & -0.076, $p=0.61$ \\ 
L$_{\textrm{1.4 GHz}}$ & 0.17, $p=0.49$ & 0.22, $p=0.5$ \\ 
\end{tabular}
\label{tab:t3}
\end{table}

\begin{figure*}
\includegraphics[width=\textwidth,clip,trim=0cm 0cm 1cm 0cm]{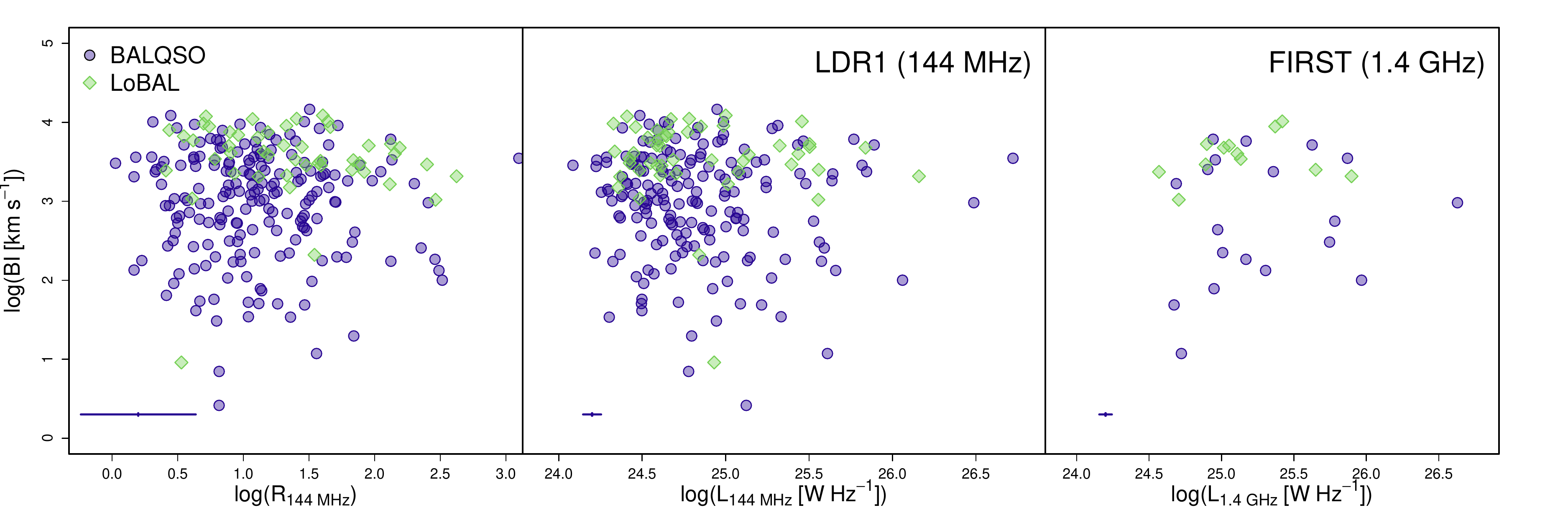}
\caption{\label{fig:f12} Balnicity index (BI) as a function of, from left to right: \logr , $L_{\textrm{144 MHz}}$, $L_{\textrm{1.4 GHz}}$. Median errors in both the x and y directions are shown in the bottom left corner of each plot.}
\end{figure*}

Finally, we investigate the dependence of the radio detection fraction of BALQSOs on BI, for both LDR1 and FIRST sources. This is shown in Figure~\ref{fig:f13}. For LDR1, the radio detection fraction increases from 0.187$^{+0.018}_{-0.017}$ to 1.000$^{+0.0}_{-0.667}$. For FIRST, the radio detection fraction increases from 0.024$^{+0.006}_{-0.006}$ to 0.087$^{+0.067}_{-0.056}$. Although the uncertainties are large for the FIRST points, and for the higher bins of the LDR1 points, the lowest and highest bins are inconsistent with each other, for both samples, indicating a positive correlation of radio detection fraction with BI. The radio detection fraction is higher for LDR1 than FIRST, but the radio detection fractions increase in a similar way. To show this, we plot the ratio of the radio detection fractions in the bottom panel of Figure~\ref{fig:f13}. This ratio is constant within the uncertainties for the entire range of BI for which there is data in both surveys. We remind the reader that we checked that these results are robust when removing FIRST-selected quasars/BALQSOs, which could bias the radio detection fraction. The correlation between radio detection fraction and BI is evidence for a physical link between the two phenomena. However, the lack of any correlation between BI and radio luminosity or radio loudness, coupled with this correlation between radio detection fraction and BI, indicates that although the radio and broad absorption lines are related to the same underlying physical process(es), they are spatially unrelated. This is inconsistent with models where the radio emission is generated directly from disc winds.

\begin{figure}
\begin{center}
\includegraphics[width=0.5\textwidth]{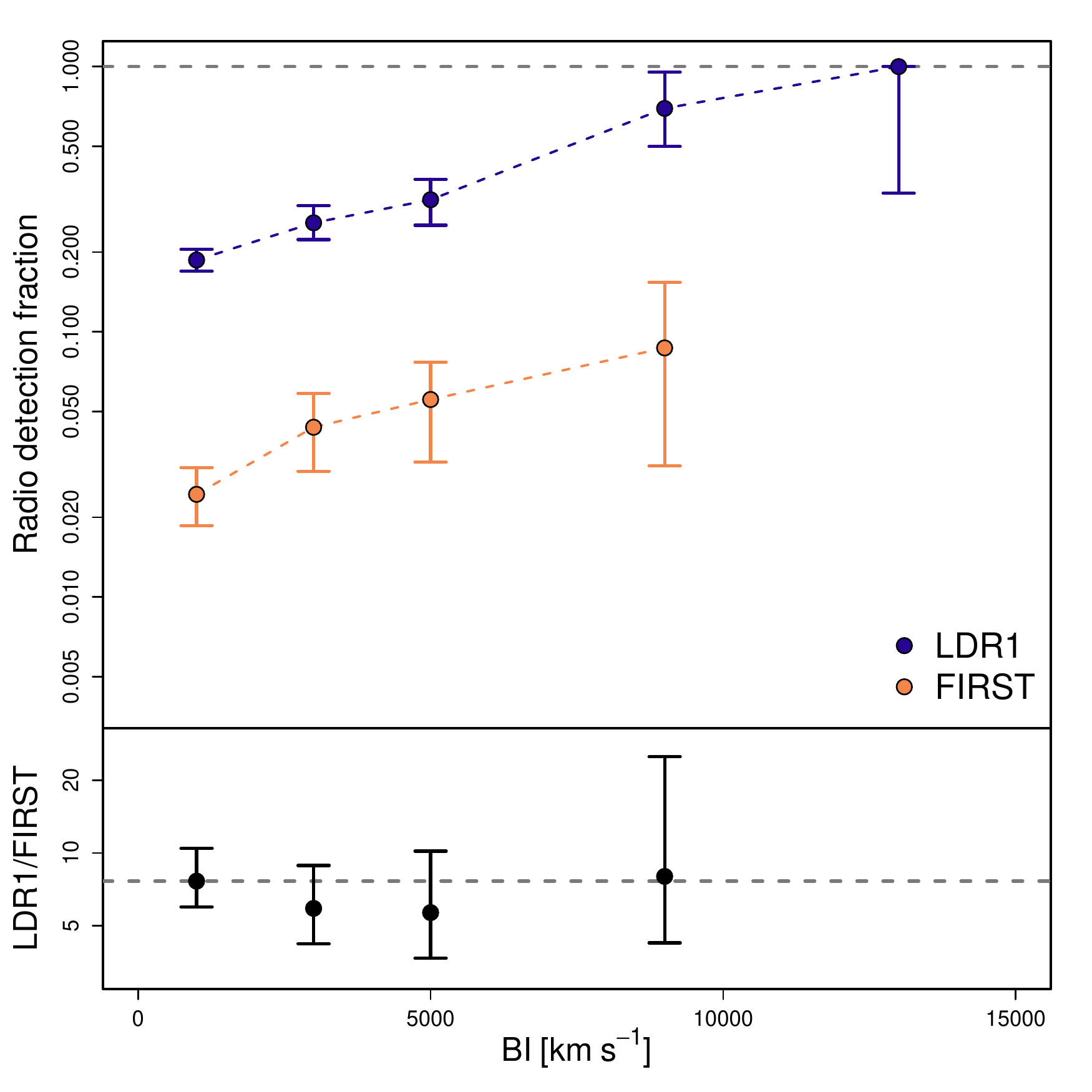}
\caption{\label{fig:f13} \textit{Top panel:} Radio detection fractions of BALQSOs as a function of balnicity index (BI), for LoTSS and FIRST (blue and orange points, respectively). A value of unity is plotted as a dashed gray line to guide the eye. \textit{Bottom panel:} Ratio of the radio detection fractions (LoTSS/FIRST), with the median value plotted as a dashed gray line. }
\end{center}
\end{figure}

\section{Discussion}
\label{sec:discussion}
The low frequency radio properties of BALQSOs studied here have implications for the origin of the BAL phenomenon, the origin of the radio emission in BALQSOs and RQQs and the general connection between accretion and outflow in quasars. 

\subsection{Where does BALQSO radio emission come from?}
Overall, our results can be explained by a scenario in which the low-frequency radio emission in BALQSOs comes from jets/lobes while the higher frequency radio emission comes from an AGN core. This is supported by the larger sizes of the radio emission at low radio frequencies when compared with high frequencies. The radio luminosities at low and high frequencies are correlated with each other, which is consistent with both processes being driven by the same AGN. One possibility is that these AGN could be undergoing rapid bursts of activity, and the low-frequency observations trace the steep-spectrum remnants of previous outbursts, while the high frequency observations trace the current AGN activity, as suggested by \cite{bruni_restarting_2015} and \cite{kunert_vlbi_2018}.

We favour the AGN scenario over radio emission from star formation or a BAL outflow itself for the following reasons. Star formation produces both free-free emission from \ion{H}{ii} regions as well as synchrotron radiation from supernova remnants. Observations at low-frequency will measure the synchrotron emission, while 1.4$\,$GHz will measure a mixture of synchrotron and free-free emission. The resolution element of FIRST is $\gtrsim 5$\sarc\ which is much larger than the typical size of galaxies at the redshifts of our sample ($z>1.7$), so we do not expect star formation to be resolved out if it is above the flux limit. For the higher frequency radio emission to be consistently coming from a smaller region than the low-frequency radio emission would require supernova remnants to be spread throughout a galaxy, and strong free-free emission from a central region. This could be explained by a central starburst following relatively recent wide-spread star formation, but this would be required in a majority of BALQSOs and is not supported by any other evidence \citep[e.g.,][]{gurkan_lofar/h-atlas:_2018}. In particular, the low-frequency radio luminosities of the BALQSO sample here are above $10^{24}\,$\wphz\ which \cite{gurkan_lofar/h-atlas:_2018} find to be too luminous for star-forming galaxies drawn from LDR1.

Although there is an at least an indirect connection between BAL outflows and their 144$\,$MHz radio emission, as evidenced by the increased LDR1 detection fraction at high BI, the lack of correlation between BI and radio power implies that BAL winds themselves do not produce the radio emission. This supports a scenario where synchrotron-emitting regions are physically distinct phenomena from BAL winds (the geometry of collimated jets make this likely). However, positing that the BAL winds do not produce the radio emission relies on BI roughly tracing the expected radio luminosity of a disc wind through its dependence on mass-loss rate. BI can also be expected to vary with the ionizing spectrum and viewing angle \citep{richards_unification_2011,higginbottom_simple_2013}. BAL trough variability \citep[e.g.][]{capellupo_variability_2011,mcgraw_quasar_2018}, due to, e.g., breaking of azimuthal symmetry \citep{dyda_non-axisymmetric_2018}, could lead to substantial scatter in BI for a constant wind kinetic power. If the radio emission originates from a larger-scale blast wave shock driven by intermittent episodes of wind activity it is plausible that BI would not correlate with radio emission, but such a model has too many uncertainties and degrees of freedom to be tested reliably. The distinction between `wind' and `jet' is also less well-defined in this case. It has been suggested that BAL outflows can lie at large distances from the central BH \citep{arav_evidence_2018,xu_vlt/x-shooter_2018}. This can be explained by a model in which the absorption troughs are formed in shocks when a quasar blast wave collides with a dense interstellar clump \citep{faucher-giguere_physical_2012}. Sources with BAL trough variability would prove particularly interesting to study at low frequencies and may permit tests of wind feedback models \citep[e.g.][]{silk_quasars_1998,king_black_2003,faucher-giguere_physics_2012,costa_feedback_2014}. Overall, a scenario where the low-frequency radio emission is due to jets is more likely than wind-driven blast waves, although these cannot be ruled out at this point. 

Our interpretation of the low-frequency radio emission stemming from jets is also supported by VLBI observations of radio-loud BALQSOs at higher frequencies, which find evidence for small-scale jets. Regardless of the morphology revealed by VLBI, BALQSOs generally have compact radio sizes \citep{doi_multifrequency_2013,kunert-bajraszewska_vlbi_2015}, and are preferentially radio-quiet \citep[][this work]{stocke_radio_1992,becker_properties_2000}. If small-scale jets are ubiquitous in BALQSOs, why do they not grow to the same sizes as radio-loud Fanaroff-Riley \citep{fanaroff_morphology_1974} type sources? It is tempting to draw the comparison to CSS and GPS sources, which are thought to be either young sources where the jets have not yet had time to grow, or sources with dense galactic environments which frustrate the jets and keep them contained on sub-galaxy scales.

\subsection{Implications for the BAL phenomenon: orientation versus evolution}
Drawing conclusions about the impact on the orientation- and evolution-based models for the BAL phenomenon is difficult. What we can say is that the physical picture we think is most likely from our results is that the radio emission arises from jets which are physically distinct from BAL winds. The tentative anti-correlation of BI with LDR1/FIRST LLS ratio suggests that the larger this ratio is (and thus more in the plane of the sky) the weaker the BAL winds are. In an orientation-only model, this would imply that the BAL winds are co-oriented with the direction of the radio jets, which we do not think is likely \citep[although polar winds have been seen in a handful of BALQSOs, see e.g.,][]{ghosh_physical_2007,zhou_polar_2006}.     This is suggestive that BALQSOs could be at a particular evolutionary stage. In the evolutionary picture, the tentative anti-correlation of LDR1/FIRST LLS ratio with BI could imply larger jets/lobes (and therefore perhaps older) are associated with weaker BALs. This is consistent with an evolutionary picture where the central quasar produces radio jets, which begin to drive the outflows we see in BALQSOs -- as the jets increase in age and clear out more material, the covering fraction of absorbing material, and thus the BI, will decrease. 

While radio spectral information can provide a proxy for orientation, the spectral indices presented here are calculated point-to-point from 144$\,$MHz to 1.4$\,$GHz and we cannot distinguish if objects have truly flat spectra or are peaked in between these two frequencies. Consequently we do not draw any conclusions from the current spectral information. 

Although our results do not clearly favour orientation or evolution dependent models, we have learned something about the geometry of the individual components of BALQSOs: the BAL winds and source of radio emission appear to be spatially distinct phenomena. As LoTSS continues to survey the Northern sky the data collected will help readdress this question with more concrete evidence. 

\subsection{Accretion properties}
\label{subsec8}
Accretion is likely to be the ultimate energy source for jets and winds across the mass scale, and there is an intimate connection between the accretion onto the object and the outflows from it; in X-ray binaries, winds and jets tend to appear in specific accretion states \citep{fender_towards_2004,ponti_ubiquitous_2012}, a  phenomenon which is also observed in accreting white dwarfs \citep{kording_detection_2011}. The latter are particularly relevant systems to quasars, as their discs peak in the UV \citep{warner_cataclysmic_1995}, they have UV-absorbing disc winds \citep{cordova_high-velocity_1982} and also emit significantly at radio wavelengths \citep{coppejans_novalike_2015}. In AGN and quasars, the connection between discs and jets is less well understood, but the principles from XRBs have been extended to both radio-loud and radio-quiet systems \citep{maccarone_connection_2003,kording_accretion_2006}.

A useful way to parameterise the accretion state of a disc is through the Eddington ratio, defined as
\begin{equation}
\eta_{\mathrm{Edd}} = 
\frac{\sigma_T}{4 \pi G m_p c} 
\frac{L_\mathrm{bol}}{M_{BH}}
\end{equation}
where $\sigma_T$ is the Thomson cross-section, $m_p$ is the proton mass, $G$ is the gravitational constant and $c$ is the speed of light. Quasar winds should care about this ratio. The discovery of line-locking signatures \citep{arav_radiative_1995,arav_ghost_1996,north_new_2006,bowler_line-driven_2014} suggests that BALQSO winds are at least partially driven by radiation pressure on spectral lines \citep[`line-driving';][]{castor_radiation-driven_1975,proga_dynamics_2000,proga_dynamics_2004}. If so, winds should be produced at relatively high Eddington ratios, when the disc is radiatively efficient and there is plenty of UV radiation to impart momentum to the flow. Given the expected connection between winds, jets, and the accretion state, it is useful to investigate the Eddington ratio distributions of our sources.

In Figure~\ref{fig:f14} we show the normalised distributions of Eddington ratios for non-BAL quasars, BALQSOs, and LoBALs with and without LDR1 detections. The Eddington ratio can be affected by large systematic uncertainties in estimates of $L_\mathrm{bol}$ \citep{richards_spectral_2006,runnoe_updating_2012} and $M_{BH}$ \citep{jarvis_orientation_2006,lamastra_possible_2006,marziani_estimating_2012,denney_are_2012,coatman_c_2016}, but it nonetheless gives us a convenient, approximate way of quantifying the accretion rate normalised to the BH mass. We find that our radio-detected sources lie almost exclusively in the range $0.01 \lesssim \eta_{\mathrm{Edd}} \lesssim 1$, roughly as expected for an optically thick, radiatively efficient accretion disc \citep[e.g.][]{shakura_black_1973,maccarone_x-ray_2003,qiao_dependence_2009}. This distribution is largely a result of the underlying distribution in the quasar catalogue. The Kolmogorov-Smirnov test shows no statistically significant differences between any of the different sub-samples of BALQSOs. The BALQSOs and LoBALs show a similar distribution of Eddington ratios to non-BAL quasars, confirming that the situation in quasars is less clear-cut than in X-ray binaries, in agreement with, e.g., \cite{sikora_radio_2007}.

\begin{figure*}
\includegraphics[width=\textwidth]{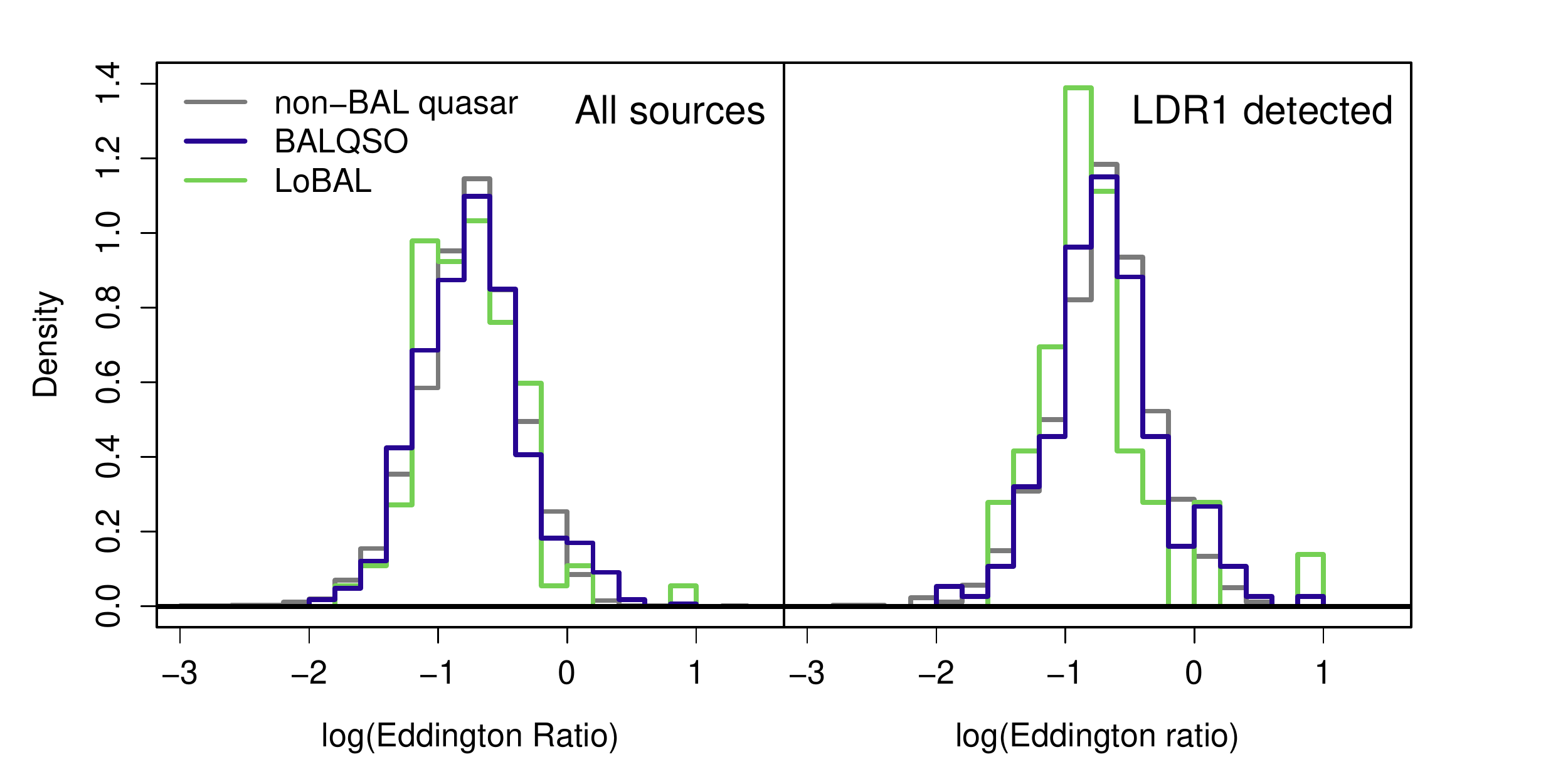}
\caption{\label{fig:f14} Normalised distributions of Eddington ratios for non-BAL quasars (grey), BALQSOs (blue) and LoBALs (green). {\em Left:} All sources -- {\em Right:} LDR1 detected sources only. }
\end{figure*}

\section{Summary and conclusions}
\label{sec:conclusions}
In this paper, we have investigated the low-frequency radio properties of BALQSOs in LDR1. We examined the radio detection fractions of BALQSOs, which we also divided into LoBALs and HiBALs. We investigated the properties of radio-loudness, radio sizes, and radio spectra. We also studied the fractions of BALQSOs/non-BAL quasars and LoBALs/BALQSOs, as a function of radio power. We were able to expand on previous studies at 1.4$\,$GHz with improved cross-matching between FIRST and SDSS via LOFAR/PS1 cross-matching. 

Our main results are as follows:

\begin{itemize}
\item BALQSOs are twice as likely to be detected than non-BAL quasars in LDR1 at 144$\,$MHz, with LoBALs having a radio detection fraction 1.6 times that of than HiBALs. This trend persists even for a sub-sample of HiBALs  with the same median BI of the LoBALs.

\item Within the subset of LDR1-detected quasars, the LoBAL, HiBAL, and overall BALQSO fractions are constant with increasing radio luminosity at 144$\,$MHz. This trend holds even when selecting only sources with radio counterparts detected in FIRST. This is inconsistent with what has previously been reported at 1.4$\,$GHz, which implies that the low and high frequencies may be tracing different sources of radio emission. 

\item The majority of BALQSOs would be classified as `radio-quiet' based on the classical definition. We do not find clear evidence of any bi-modality of `radio-quiet' and `radio-loud' BALQSOs, although a slight bi-modality does appear in the distribution of \logr\ for non-BAL quasars. 

\item The radio sizes of BALQSOs at 144$\,$MHz are generally less than about 200 kiloparsecs. When comparing to FIRST radio sizes, we find that BALQSOs tend to be larger at 144$\,$MHz than at 1.4$\,$GHz, consistent with systems dominated by jets/lobes at low frequencies and AGN cores at high frequencies.

\item The radio spectral indices of BALQSOs, in particular LoBALs, between 144$\,$MHz and 1.4$\,$GHz tend to be flatter than those of non-BAL quasars, although whether this is due to intrinsically flat radio spectra or radio spectra peaking in between these two frequencies is unclear. 

\item The fraction of BALQSOs which are LoBALs remains constant within the uncertainties with increasing radio luminosity at 144$\,$MHz. 

\item We find no correlation between BI and \logr , \llofar , or \lfirst . We do find that the radio detection fraction in both LDR1 and FIRST increases with increasing BI, and that this happens in the same way for both surveys. 

\item The fact that the radio detection fraction is correlated with BI, but {\em not} radio properties such as luminosity or radio-loudness, indicates that the radio emission and BI are initiated by the same process, but are physically separated from each other. That is, the radio emission cannot be generated by the same disc winds that drive the BI. 

\end{itemize}

In the future, LoTSS will cover the entire Northern sky, providing tens of thousands of radio-detected BALQSOs. Such a large sample will enable us to improve on the work presented here by reducing the uncertainties and allowing us to refine bins in radio power and BI to further investigate the dependence of radio-detected BALQSOs on these properties. In the meantime, follow up studies of this sample at intermediate frequencies (e.g, the Giant Metre-wave Radio Telescope at 610 MHz) will help determine the shape of the radio spectra of BALQSOs, which we were not able to do here. Future data releases from LoTSS will include in-band spectral indices, which will provide further information. Finally, by using the international stations of LOFAR we can achieve sub-arcsecond resolution to observe the spatially resolved low-frequency morphology of LoTSS-detected BALQSOs.

\section*{Acknowledgements}
LKM acknowledges financial support from Oxford Hintze Centre for Astrophysical Surveys which is funded through generous support from the Hintze Family Charitable Foundation. This publication arises from research partly funded by the John Fell Oxford University Press (OUP) Research Fund. JHM acknowledges financial support from STFC grant ST/N000919/1. WLW acknowledges support from the UK Science and Technology Facilities Council [ST/M001008/1]. PNB and JS are grateful for support from the UK STFC via grant ST/M001229/1. GG acknowledges the CSIRO OCE Postdoctoral Fellowship. IP acknowledges support from INAF under PRIN SKA/CTA ‘FORECaST’. KJD and HJAR acknowledge the support from the European Research Council under the European Unions Seventh Framework Programme (FP/2007- 2013) /ERC Advanced Grant NEWCLUSTERS-321271. MJH acknowledges support from the UK Science and Technology Facilities Council [ST/M001008/1]. MKB acknowledges support from the National Science Centre (Poland) under grant no. 2017/26/E/ST9/00216. SM acknowledges funding through the Irish Research Council New Foundations scheme and the Irish Research Council Postgraduate Scholarship scheme. 

This paper is based (in part) on data obtained with the International LOFAR Telescope (ILT) as part of project code LC2\_038 and LC3\_008. LOFAR (van Haarlem et al. 2013) is the Low Frequency Array designed and constructed by ASTRON. It has observing, data processing, and data storage facilities in several countries, that are owned by various parties (each with their own funding sources), and that are collectively operated by the ILT foundation under a joint scientific policy. The ILT resources have benefitted from the following recent major funding sources: CNRS-INSU, Observatoire de Paris and Université d'Orléans, France; BMBF, MIWF-NRW, MPG, Germany; Science Foundation Ireland (SFI), Department of Business, Enterprise and Innovation (DBEI), Ireland; NWO, The Netherlands; The Science and Technology Facilities Council, UK; Ministry of Science and Higher Education, Poland.

The data used in work was in part processed on the Dutch national e-infrastructure with the support of SURF Cooperative through grant e-infra 160022 \& 160152. This research has made use of data analysed using the University of Hertfordshire high-performance computing facility (\url{http://uhhpc.herts.ac.uk/}) and the LOFAR-UK computing facility located at the University of Hertfordshire and supported by STFC [ST/P000096/1]. 

This research has made use of \textquotedblleft Aladin sky atlas\textquotedblright\ developed at CDS, Strasbourg Observatory, France. Funding for the Sloan Digital Sky Survey has been provided by the Alfred P. Sloan Foundation, the U.S. Department of Energy Office of Science, and the Participating Institutions. SDSS acknowledges support and resources from the Center for High-Performance Computing at the University of Utah. The SDSS web site is www.sdss.org. This research made use of: Astropy, a community-developed core Python package for Astronomy \citep{astropy_2013,astropy_2018}; APLpy, an open-source plotting package for Python \citep{aplpy_2012} and astroML \citep{astroML_2012}. 

\clearpage 
\clearpage 
\clearpage 
\bibliographystyle{aa}
\bibliography{ms.bib}

\appendix
\section{A selection of images}
\label{appendix1}
All image cutouts are 100\sarc\ per side, and the insets are 10\sarc\ per side. White contours show LoTSS emission while red contours show FIRST emission. The first two contours are always at 3 and 5 times the median absolute deviation (MAD) of pixel values in the cutout images, and three more contours are evenly spaced between 5 times the MAD and 0.8$\times$maximum value in the image. The background images are RGB made from SDSS $g,r,i$ bands. The LDR1 beam is shown in the bottom right hand corner. The beam parameters are \textsc{bmaj=}6\sarc\ , \textsc{bmin=}6\sarc\ , and \textsc{bpa=}90 degrees. The FIRST beam (not pictured) has similar parameters, with \textsc{bmaj=bmin=}5.4\sarc\ and \textsc{bpa=}0 degrees.

\begin{figure*}
\begin{center}
\includegraphics[width=\textwidth,clip,trim=0cm 19.75cm 0cm 0cm]{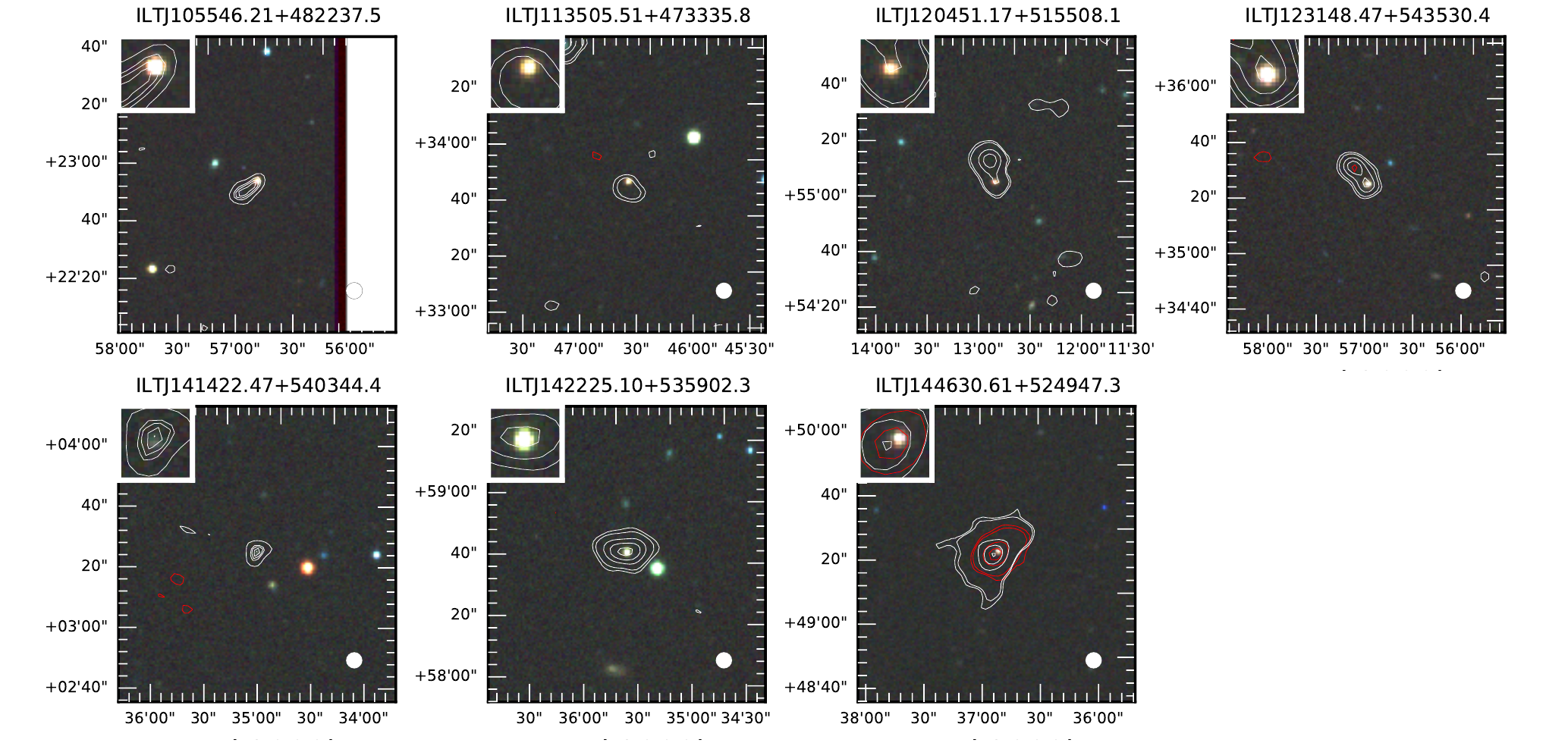}
\caption{BALQSOs which are resolved in LDR1. J141422.47+540344.4 is a LoBAL and the rest are HiBALs.}
\end{center}
\end{figure*}

\begin{figure*}
\begin{center}
\includegraphics[width=\textwidth,clip,trim=0cm 14.5cm 0cm 0cm]{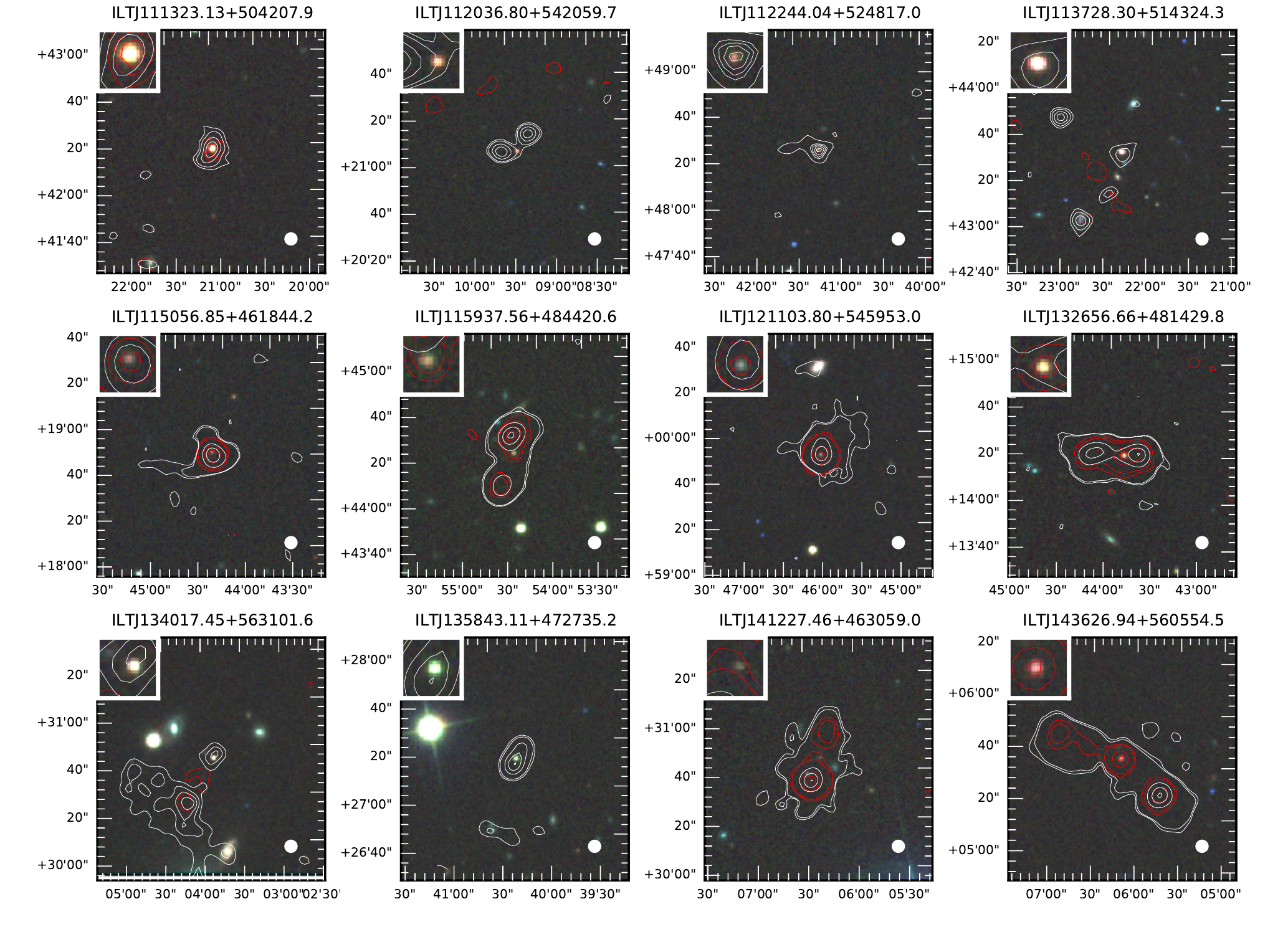}
\caption{\label{fA:f2} A selection of non-BAL quasars which are resolved in LDR1.}
\end{center}
\end{figure*}

\begin{figure*}
\begin{center}
\includegraphics[width=\textwidth,clip,trim=0cm 19.85cm 0cm 0cm]{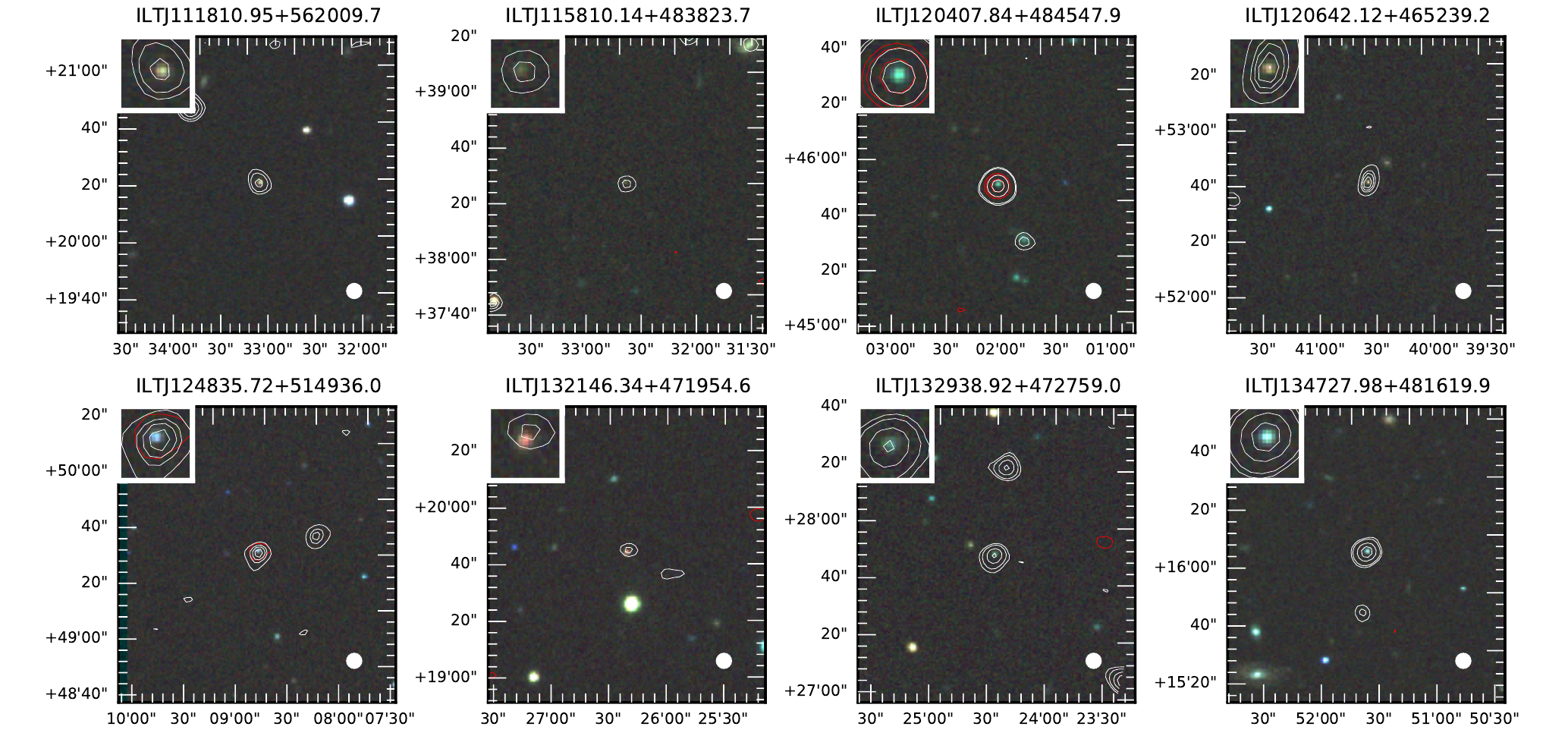}
\includegraphics[width=\textwidth,clip,trim=0cm 24.5cm 0cm 0cm]{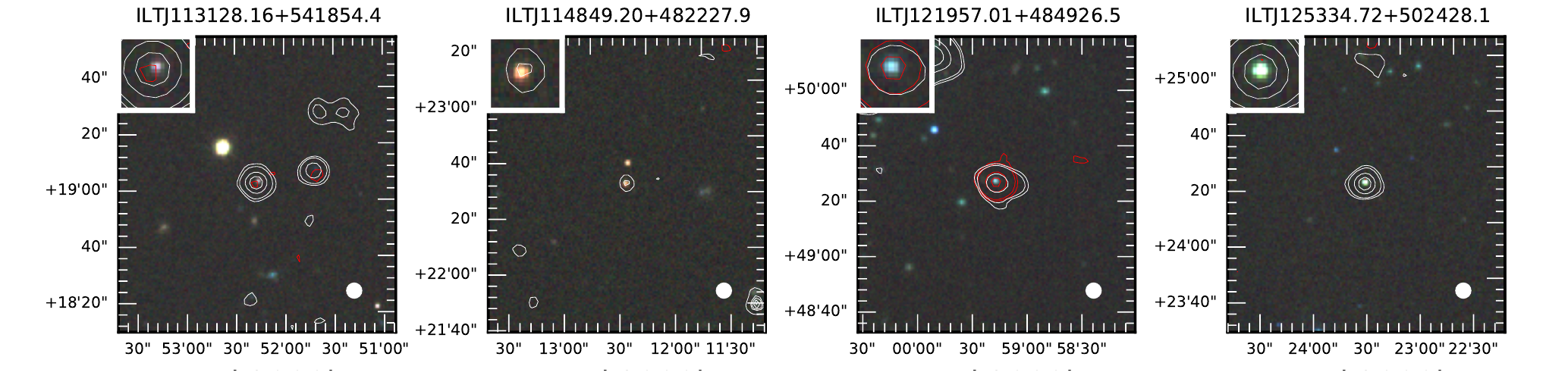}
\caption{\label{fA:f3} A selection of unresolved LoBALs (top two rows) and HiBALs (bottom row).}
\end{center}
\end{figure*}

\begin{figure*}
\begin{center}
\includegraphics[width=\textwidth,clip,trim=0cm 19.75cm 0cm 0cm]{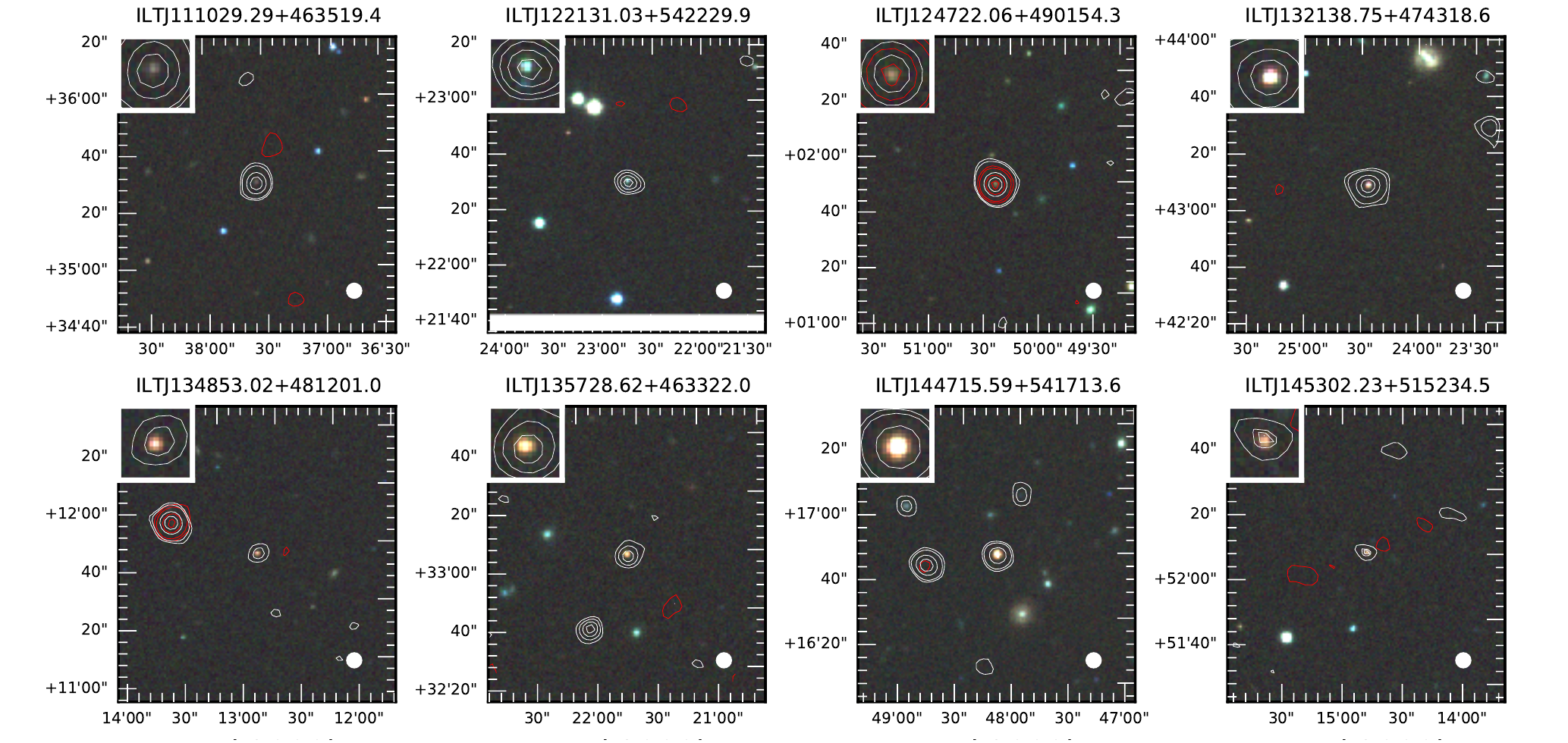}
\caption{\label{fA:f4} A selection of unresolved  non-BAL quasars.}
\end{center}
\end{figure*}

\end{document}